\documentclass{egpubl}
\usepackage{eg2027}

\ConferenceSubmission

\usepackage[T1]{fontenc}
\usepackage{dfadobe}

\usepackage{tikz}
\usepackage[ruled]{algorithm2e}

\SetAlFnt{\small}
\SetAlCapFnt{\small}
\SetAlCapNameFnt{\small}
\SetAlCapHSkip{0pt}

\usepackage{multirow} 
\usepackage{array}
\usepackage{xcolor}
\usepackage{overpic}
\usepackage{booktabs}
\newcommand{\FLIP}{\reflectbox{F}LIP}

\newcommand{\figref}[1]{Figure~\ref{#1}}
\newcommand{\secref}[1]{Section~\ref{#1}}

\newcommand{\tabref}[1]{Table~\ref{#1}}

\usepackage{subfigure}
\usepackage{caption}
\usepackage{amsmath}

\BibtexOrBiblatex

\electronicVersion
\PrintedOrElectronic

\usepackage{graphicx}
\usepackage{egweblnk}

\title[Real-time Rendering of Pre-integrated Neural Emitters]{Real-time Rendering of Pre-integrated Neural Emitters}

\author[Arno Coomans, Floor Verhoeven, Edoardo Dominici, Markus Steinberger]
{\parbox{\textwidth}{\centering
Arno Coomans$^{1}$\orcid{0000-0002-7396-5044} \ \
Floor Verhoeven$^{2}$\orcid{0000-0003-3768-0460} \ \
Edoardo A. Dominici$^{1,2}$\orcid{0009-0007-9101-7279} \ \
Markus Steinberger$^{1,3}$\orcid{0000-0001-5977-8536}
        }
        \\
{\parbox{\textwidth}{\centering 
Graz University of Technology, Austria$^1$ \\
Huawei Technologies, Switzerland$^2$, Austria$^3$}
}
}

\begin{document}

\teaser{
\newlength{\teaserImgWidth}
\setlength{\teaserImgWidth}{0.156\textwidth}
\newlength{\teaserImgHeight}
\setlength{\teaserImgHeight}{0.22\linewidth}
\newcommand{\horSpace}{\hspace{1.2pt}}
\newcommand{\lastSpace}{\hspace{1.5pt}}
\newcommand{\verSpace}{2.5pt}
\captionsetup[subfigure]{labelformat=empty}
\begin{center}
\makebox[\linewidth][s]{%
 \includegraphics[width=0.3275\linewidth]{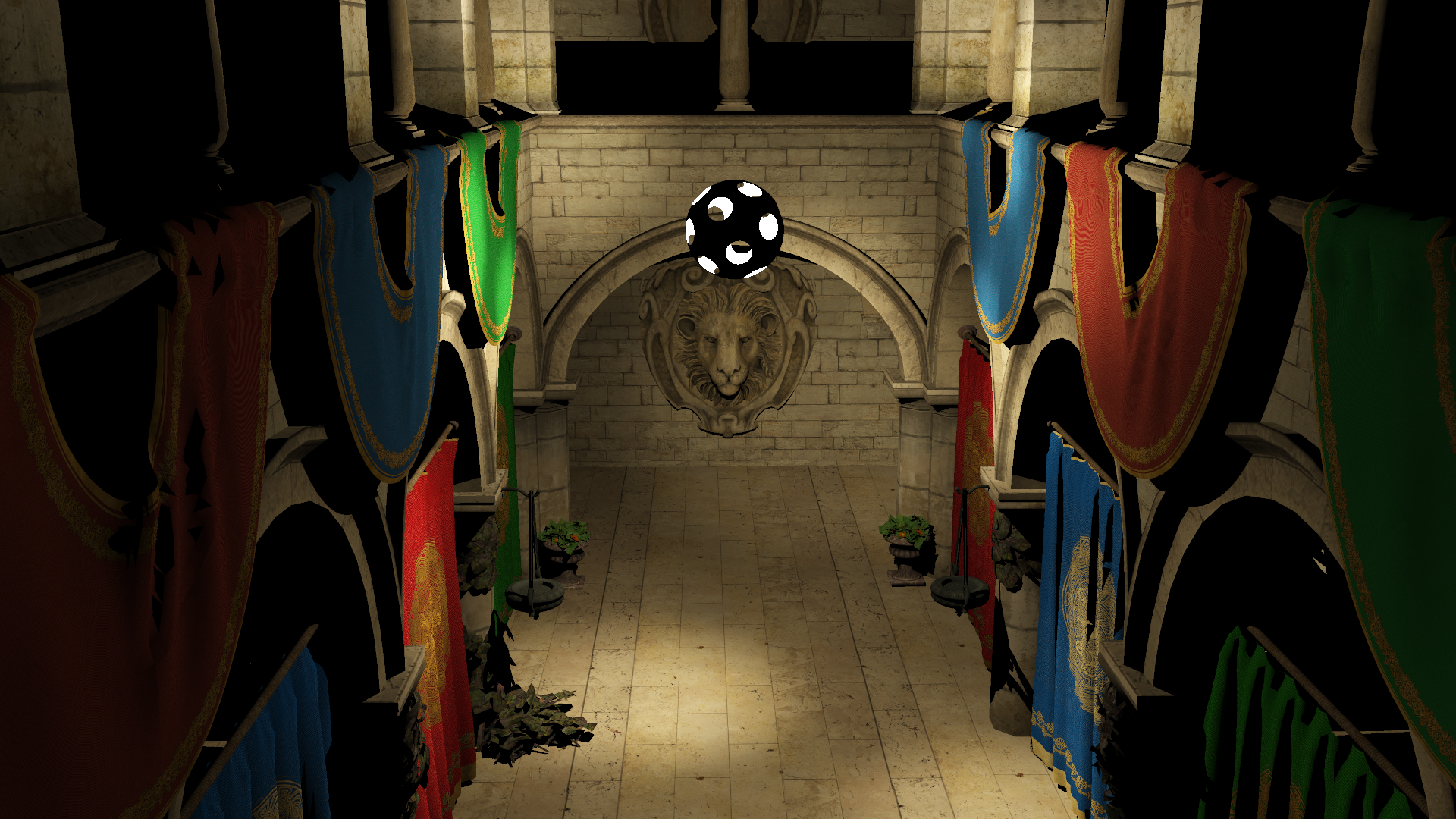}\hfill
 \includegraphics[width=0.3275\linewidth]{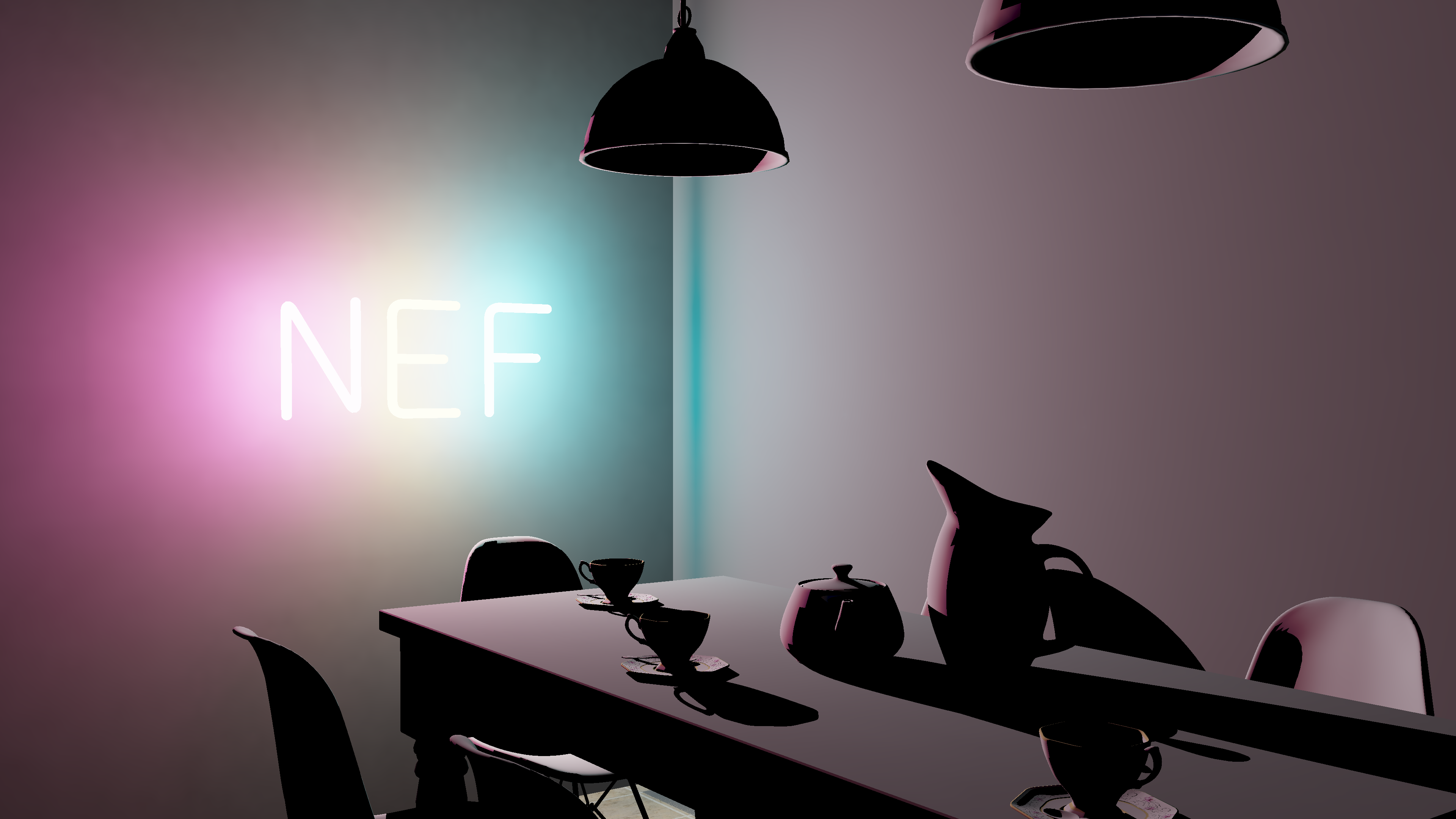}\hfill
 \includegraphics[width=0.3275\linewidth]{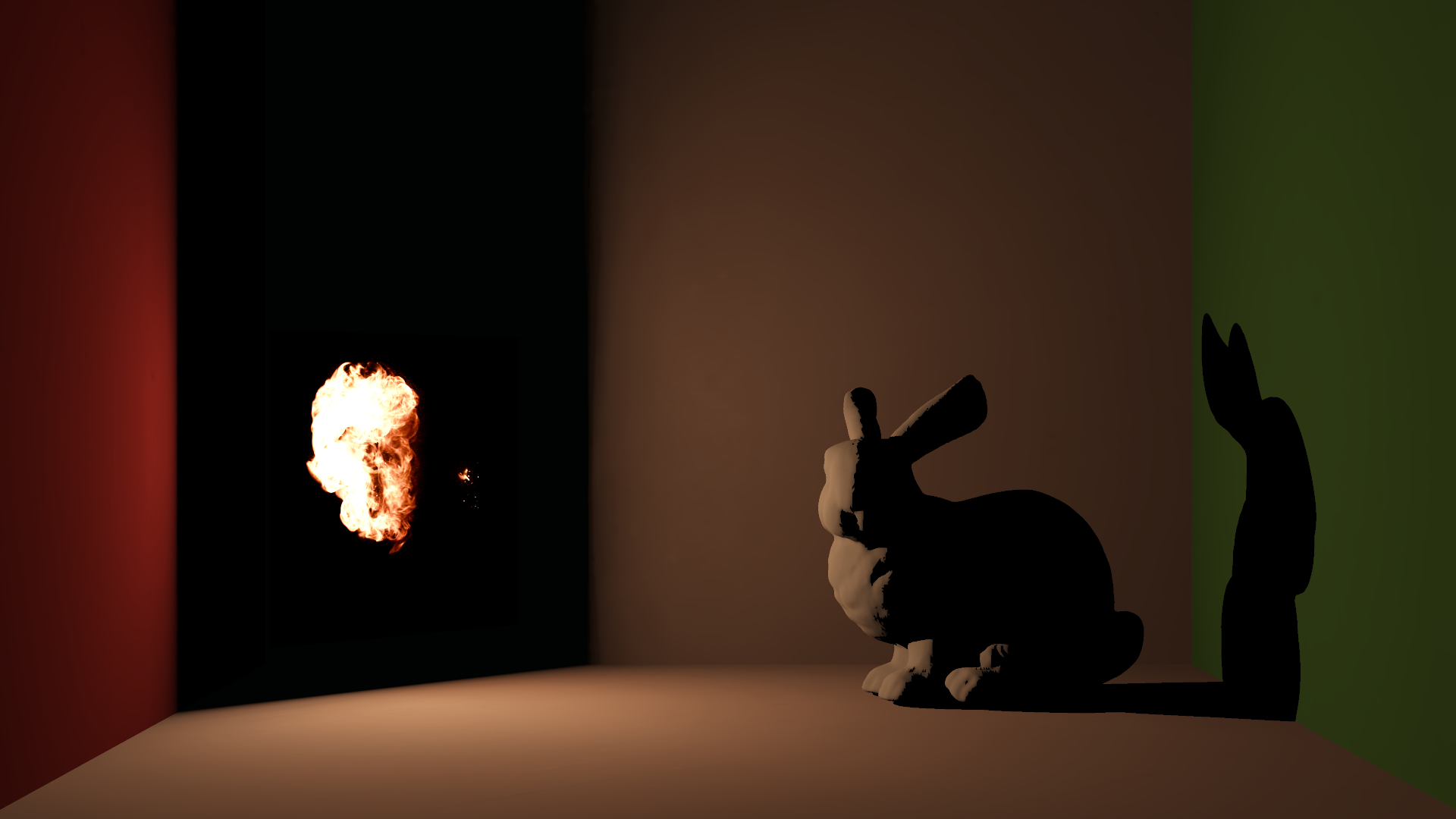}%
}
\end{center}
\centering
\caption{A Neural Emission Field (NEF) pre-integrates illumination around non-trivial emitters into a neural representation that can be placed in any scene after training. From left to right: a complex emitter with pre-integrated interreflections; a spatially varying emission profile; and a dynamic fireball whose emission changes over time. Querying the NEF in full-HD takes 1.2 ms for the above scenes.
}
\label{fig:teaser}
}

\maketitle
\thispagestyle{plain}
\makeatletter
\def\@oddhead{}\def\@evenhead{}
\def\@oddfoot{\hfil\thepage\hfil}\def\@evenfoot{\hfil\thepage\hfil}
\makeatother
\begin{abstract}
Direct illumination from non-trivial emitters with occluding housing, high-polygon emissive meshes, spatially varying emission, or deformable assemblies is a persistent bottleneck in real-time rendering.
Current real-time solutions either rely on simplified analytical representations or fall back to runtime sampling.
Analytical methods are restricted to simple emitter geometry, pure sampling-based estimators need high sampling budgets to be noise-free, and proxy representations still require runtime integration over outgoing radiance around the emitter.
We present Neural Emission Fields (NEF), a representation that eliminates runtime integration by precomputing the illumination in the volume around the emitter. This neural field is parameterized by position, normal, view direction, and material parameters. Using a two-headed diffuse/glossy architecture, it yields noise-free unoccluded direct illumination from a single network evaluation per shading point. Because training is performed in the emitter's local frame, a trained NEF acts as a portable lighting asset reusable across scenes under rigid transforms. Internal interreflections, self-occlusion, spatially varying emission, and deformations are absorbed into the learned representation at zero additional runtime cost. 


\end{abstract}

\section{Introduction}
Direct illumination is a major cost in real-time rendering pipelines, especially when emitters are non-trivial---luminaires with housing geometry that occludes the internal filament, emissive meshes with thousands of polygons, light sources with spatially varying or textured emission profiles, or deformable emitters that change shape over time. While path tracing~\cite{kajiya1986rendering} handles all of these correctly, converging to noise-free direct illumination from non-trivial emitters is incompatible with real-time rendering budgets. Existing alternatives that do meet those rendering budgets typically compromise on generality, accuracy, or noise-free output. 

Analytical methods like Linearly Transformed Cosines (LTC) \cite{ltc2016} are extremely fast but assume 2D polygonal emitters and scale linearly with the emitter's vertex count. For more general emitters, Monte Carlo estimators of direct illumination, including reservoir methods~\cite{bitterli2020spatiotemporal}, support arbitrary emitters but produce noise that must be removed by a denoiser, which often introduces unwanted temporal and spatial artifacts. Proxy-based representations~\cite{heidrich1998canned, velazquez2015complex}, including recent neural variants~\cite{zhu2021neural}, sidestep both limitations by precomputing the outgoing radiance into a 4D near-field proxy, factoring out the cost of sampling paths within the emitter assembly. However, receiver points must still integrate over this proxy at runtime.

Our key insight is that this remaining runtime integration is avoidable. Rather than precomputing only the radiance \emph{leaving} an emitter (which still needs to be integrated), we precompute the pre-integrated outgoing direct illumination with respect to a receiver point (any shaded surface point in the volume around the emitter). We represent this as a neural field parameterized by the receiver's position, normal, view direction, and material parameters, learned by a neural network conditioned on a material model. By shifting the integration to a one-time precomputation phase, a single network evaluation replaces the entire integral over the (proxy) emitter. 

Because the network is trained in the emitter's local coordinate frame, the resulting model functions as a portable lighting asset. World-space queries are mapped into the emitter's frame at shading time rather than baked into the network weights, so a trained NEF can be translated, rotated, and uniformly scaled at runtime without retraining. A single training run therefore amortizes across every scene and every instance in which the emitter is placed---much like a baked texture or precomputed BRDF. 

This design facilitates several critical performance advantages over traditional methods. Inference cost is fundamentally decoupled from the geometric complexity of the emitter; the computational cost is determined by the network's size rather than the polygon count of the emitter.
Furthermore, because the receiver's BRDF is incorporated into the learning objective, complex emissive effects are absorbed into the representation at no additional runtime cost.
Finally, since the training phase relies solely on path-traced radiance estimates for training data, the same training procedure is used for simple planar polygons, complex emissive meshes, and deformable emitter assemblies. 

We summarize our contributions as follows:
\begin{itemize}
    \item A reformulation of direct illumination as a pre-integrated, receiver-aware quantity, eliminating the runtime integration retained by prior analytical, sampling-based, and proxy-based methods.
    \item A neural emission field employing a two-headed diffuse/glossy architecture and reflected-direction encoding, yielding noise-free direct illumination and accurate specular highlights from a single network evaluation per shading point.
    \item Scene-independent training in the emitter's local coordinate frame, so a single pre-trained NEF can be translated, rotated and scaled across different scenes without retraining. 
    \item Extensions to complex emissive assemblies with internal interreflections, spatially varying emission, and deformable emitters which are non-trivial or unsupported in state-of-the-art real-time methods.
\end{itemize}

\section{Related Work}
We categorize previous work into three primary categories: analytical solutions, Monte Carlo estimators, and proxy-based representations. Finally, we discuss the broader context of using neural networks in the context of rendering.

\textbf{Analytical Solutions.}
Analytical closed-form solutions for simple emitter geometries (e.g., planar polygons) exist for various receiver types, including diffuse~\cite{baum1989} or glossy receivers like the Phong distribution~\cite{arvo1995}, and more general glossy reflections~\cite{ltc2016, kuge2022freeform, lecocq2016accurate}.
Analytical methods are extremely efficient for low-polygon emitters, but their cost scales linearly with the number of emitter vertices~\cite{arvo1995}, making them prohibitively expensive for high-polygon emitters in real-time applications. Furthermore, even on supported geometry they diverge from a path-traced reference when receivers are glossy or emission is non-uniform~\cite{ltc2016}.
Our neural approach achieves more accurate shading with respect to a ground truth solution, natively supports non-convex and non-planar emitters, and decouples cost from the emitter geometry, drastically reducing shading time for complex emissive meshes.

\textbf{Many-Light Sampling.} 
Monte Carlo integration supports estimating direct illumination from any emitter but requires specialized sampling strategies to manage runtime noise.
Early techniques addressed the many-lights problem by ranking lights according to their potential contribution \cite{ward1994adaptive} or by classifying them as important or unimportant to guide sampling \cite{shirley1996monte}. 
In complex scenes, hierarchical representations~\cite{walter2005lightcuts, moreau2019dynamic, yuksel2020stochastic} can be used to enable emitter sampling at sublinear cost. 
More recently, ReSTIR \cite{bitterli2020spatiotemporal} introduced spatiotemporal reservoir resampling, achieving high-quality direct illumination under tight sampling budgets.
Despite these advancements, the residual variance inherent to all Monte Carlo approaches typically necessitates a denoiser. In contrast, NEF avoids runtime sampling entirely, shifting the computational burden to a (one-time) per-emitter offline training phase.

\textbf{Classical Pre-integration.} 
Precomputation in rendering~\cite{ramamoorthi2009precomputation} trades offline computations for runtime cost by tabulating light transport over a discretised parameter domain.
Precomputed radiance transfer (PRT)~\cite{sloan02} projects the transfer function at each surface point onto a low-order spherical harmonics basis, reducing shading under distant lighting to a dot product. All-frequency variants require non-linear wavelet approximations to keep storage tractable~\cite{ng2003all}.
More fundamentally, PRT relies on the distant-lighting assumption, which breaks down for near-field emitters whose incident lighting varies with receiver position.

A complementary line of work stores lighting in a volume rather than on surfaces.
Irradiance volumes~\cite{greger1998irradiance} and their dynamic, ray-traced successors~\cite{majercik2019dynamic} cache irradiance at probes on a grid and interpolate between them at runtime.
For strongly spatially varying signals, such as the near field of a complex emitter, these caches require careful probe placement~\cite{silvennoinen2017real, wang2019fast} and fine discretisation to avoid interpolation and light-leaking artifacts.

These approaches share a common limitation: the storage of a tabulated representation grows exponentially with the dimensionality of the signal.
Pre-integrated direct illumination around a near-field emitter depends on receiver position (3D), normal (2D), and reflected direction (2D), which is already seven dimensions for a single glossy material.
Conditioning on roughness to cover a continuous family of materials, as done in this paper, adds an eighth parameter.
A regular grid with only $32$ samples per dimension would already require $32^8 \approx 10^{12}$ entries.
Adding a time parameter for dynamic emitters worsens this further: a naive per-keyframe cache scales linearly with the number of keyframes, and complex temporal variation demands many of them.
Like prior neural precomputation methods~\cite{rainer2022neural, coomans2026real}, NEF sidesteps this curse of dimensionality by fitting a compact, global representation that efficiently captures correlations in the high-dimensional signal.

\begin{figure*}
    \centering
    \includegraphics[width=\linewidth]{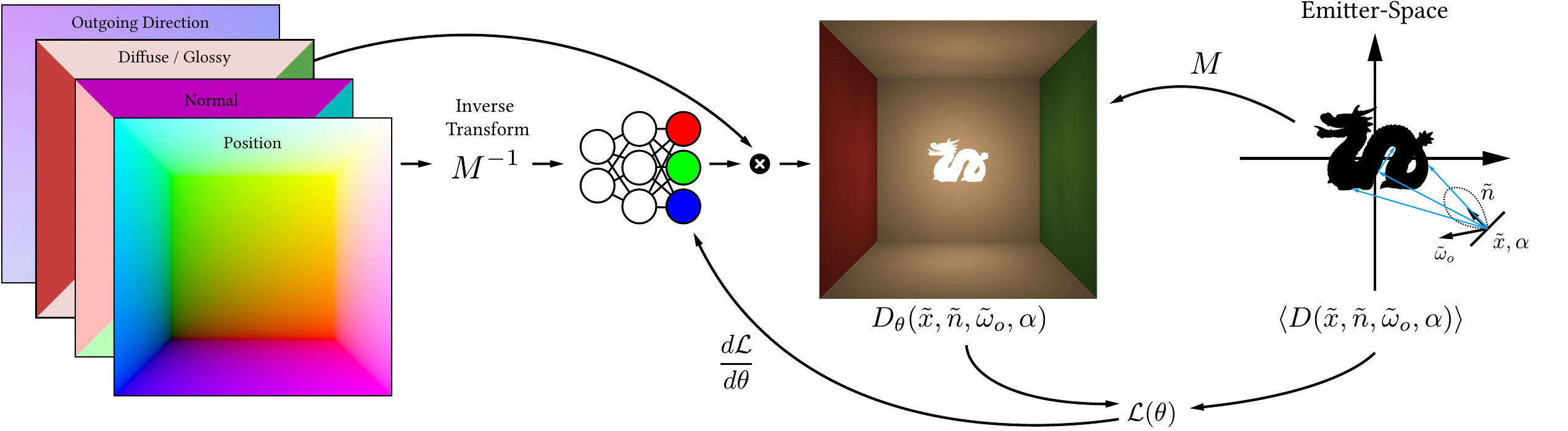}
    \caption{NEF inference and training.  \emph{Inference (left):} G-buffer channels are transformed from world space to the emitter's local frame using $M^{-1}$, fed through $D_\theta$ in a single evaluation per shading point, and modulated by the diffuse and glossy albedos to yield the final shaded image.
    \emph{Training (right):} for an emitter placed at the origin of its local frame, we sample receiver configurations $(\tilde{x}, \tilde{n}, \tilde{\omega}_o, \alpha)$ within the emitter-space bounding region and obtain target radiance values $\langle D(\tilde{x}, \tilde{n}, \tilde{\omega}_o, \alpha)\rangle$ via offline path tracing. The neural field $D_\theta$ is fit to these targets by minimising the loss $\mathcal{L}(\theta)$.
    }
    \label{fig:training_and_forward}
\end{figure*}

\textbf{Pre-integrated Emitters.}
Estimating the radiance emitted by complex light sources is non-trivial, particularly when light paths are obstructed by housing geometry or undergo multiple internal reflections.
Early research tackled this by replacing the emitter with a 4D proxy geometry. Canned light sources represent the radiance leaving the emitter's bounding surface as a 4D light field slab so that scene points can be illuminated by ray-tracing against the proxy rather than the intricate original geometry~\cite{heidrich1998canned}. 
Later work decoupled a luminaire's dual roles in rendering: a radiance volume captures its appearance when viewed directly, while virtual point lights approximate how it illuminates the surrounding scene~\cite{velazquez2015complex}. Neural variants \cite{zhu2021neural} reduced the storage requirements of these tabulated representations and combined it with a learned importance sampling distribution. Subsequent work learned radiance-field representations of the luminaire for faster evaluation~\cite{condor22}, or relied on wavelet-compressed light fields for fast per-ray reconstruction and hierarchical importance sampling~\cite{atanasov2025wavelet}.

Despite these advances, these methods still treat the proxy as an emitter that must be integrated at runtime. Each shading point integrates the product of the radiance sampled from the proxy's 4D near-field and the material evaluation.
In contrast, NEF pre-integrates this contribution offline, storing the full outgoing pre-integrated direct illumination to be queried at runtime.

\textbf{Rendering with Neural Networks.}
A common use of neural networks in rendering is capturing the radiance distribution during a training phase which amortizes over subsequent viewings of the scene. 
Early work anchored a learned radiance representation to scene geometry for interactive viewing from arbitrary directions~\cite{ren2013global}. 
Modern variants extend this idea to high-quality static scenes~\cite{hadadan2021neural}, rotatable environment lighting~\cite{rainer2022neural}, scene animations~\cite{coomans2024real}, variable scene configurations~\cite{diolatzis2022active, su2024dynamic}, and volumetric irradiance caching~\cite{coomans2026real}. These methods all rely on offline, scene-specific training. Alternatively, a small neural representation can be trained online as a radiance cache~\cite{muller2021real} with the additional runtime cost of ray tracing and denoising. 
NEF differs in that it is an offline method whose representation is anchored in the emitter's local 3D extent rather than on the 2D scene manifold, making it a portable, reusable asset across scenes.

\section{Neural Pre-integrated Emitters} \label{sec:method}

Prior proxy methods reduce the runtime cost of internal emitter transport but still leave the receiver-side integral to be evaluated per shading point. We close this remaining gap by learning the full pre-integrated direct illumination as a neural field around the emitter. Section~\ref{sec:pre_integrated} derives this formulation; Section~\ref{sec:emitter_space} defines the emitter-local frame that makes the representation portable; Section~\ref{sec:diffuse_glossy_sep} introduces the diffuse/glossy decomposition; Section~\ref{sec:train} describes training; and Section~\ref{sec:render} outlines rendering.

\subsection{Pre-integrated Emitter} \label{sec:pre_integrated}

Computing the direct illumination $L_o$ with respect to a complex emitter involves two layers of integration: the recursive internal light transport inside the emitter, and the receiver's shading integral over the hemisphere. Proxy-emitter methods~\cite{heidrich1998canned, velazquez2015complex, zhu2021neural} already peel away the inner layer by precomputing internal transport into a function $\mathcal{P}$ defined on a bounding surface.
The receiver, however, must still integrate against $\mathcal{P}$ at runtime.
We take this idea one step further by absorbing the receiver's direct illumination integral into the learned field $D_\theta$ itself:
\begin{align}
L_o(x, \omega_o) =\ & \int_\Omega \underbrace{\Bigl[\int_\Omega \cdots\, \mathrm{d}\omega_i'\Bigr]}_{\text{internal transport}} f_r(x, \omega_i, \omega_o)\, (\omega_i \cdot n)\, \mathrm{d}\omega_i \tag{reference} \\
=\ & \int_\Omega \underbrace{\mathcal{P}(t(x,\omega_i), -\omega_i)}_{\text{baked offline}} f_r(x, \omega_i, \omega_o)\, (\omega_i \cdot n)\, \mathrm{d}\omega_i \tag{proxy} \\
\approx\ & \underbrace{D_\theta(\tilde{x}, \tilde{n}, \tilde{\omega}_o, \alpha)}_{\text{baked offline}}
\tag{ours}
\end{align}
where $(x, n, \omega_o)$ defines the world-space position, normal and outgoing direction at an intersection point, 
$(\tilde{x}, \tilde{n}, \tilde{\omega}_o)$ are their counterparts expressed in the emitter's local frame (Section~\ref{sec:emitter_space}), $\omega_i$ is the incident direction,
$t(x, \omega_i)$ returns the point where the ray from $x$ in direction $\omega_i$ intersects the proxy surface, $\alpha$ are the material parameters for the BRDF $f_r$, $(\omega_i \cdot n)$ is the cosine foreshortening term, and $\Omega$ denotes the hemispherical integration domain.

Crucially, a pre-integrated quantity is not a property of the emitter alone. Since the BRDF $f_r$ appears inside the direct illumination integral, different receivers will observe different values of $D_\theta$ at the same $(x, n)$. The receiver's shading model is therefore part of what is being learned. We instantiate this using the GGX microfacet model~\cite{walter2007microfacet} over a roughness range of $\alpha \in [0.1, 1]$. 
By exposing $\alpha$ as input to the network, a single trained NEF represents the full roughness range for that material. While we restrict our scope to GGX in this work, our approach is generic and can be extended to other materials, such as Disney or layered BRDFs, by extending the conditioning inputs and the training distribution.

At runtime, evaluating $D_\theta$ requires only a single network inference at the shading point. This eliminates the need for any sampling of the proxy $\mathcal{P}$, importance sampling, and denoising.

Because $D_\theta$ is agnostic to the source of the radiance field, all the complexity that would otherwise have to be handled at runtime is shifted to the offline solver and baked into the network weights. This includes complex effects such as self-occlusion, multi-bounce transport and non-uniform emission (\secref{sec:meshes})---effects that often violate the assumptions made by analytical methods such as LTC~\cite{ltc2016}.
As a result, $D_\theta$ can stand in for a path-traced estimate of direct illumination from an emitter, provided the field can be trained to faithfully represent it.

\subsection{Emitter-space} \label{sec:emitter_space}
Our neural field is defined in the emitter's local frame (see \figref{fig:training_and_forward}). Given a rigid world transform $M$, we map a world-space shading configuration $(x, n, \omega_o)$ to its local counterpart $(\tilde{x}, \tilde{n}, \tilde{\omega}_o)$ by applying $M^{-1}$ to the position and the rotational part of $M^{-1}$ to the normal and outgoing directions.

Pre-integrated direct illumination is only physically meaningful within a bounded region around the emitter. Emitted radiance falls off with distance, and beyond a certain range an emitter's contribution becomes negligible. We therefore define the \textit{emitter-space} as an axis-aligned bounding region in the local frame, sized to extend 20$\times$ the emitter's largest extent on each axis---well beyond the 5$\times$ threshold at which the inverse-square law becomes an acceptable approximation for emitters~\cite{rea2000iesna}. Shading points inside this region are evaluated by a direct query of $D_\theta$, which is sufficient for all emitters in this work.

\begin{figure}
    \centering
    \setlength{\tabcolsep}{2pt}
    \begin{tabular}{@{}>{\centering\arraybackslash}m{0.04\linewidth} @{\hspace{4pt}} >{\centering\arraybackslash}m{0.30\linewidth} >{\centering\arraybackslash}m{0.30\linewidth} >{\centering\arraybackslash}m{0.30\linewidth}@{}}
        & {\small MC reference} & {\small NEF} & {\small NEF + fall-off} \\[2pt]
        \rotatebox{90}{\small Quad} &
        \includegraphics[width=\linewidth]{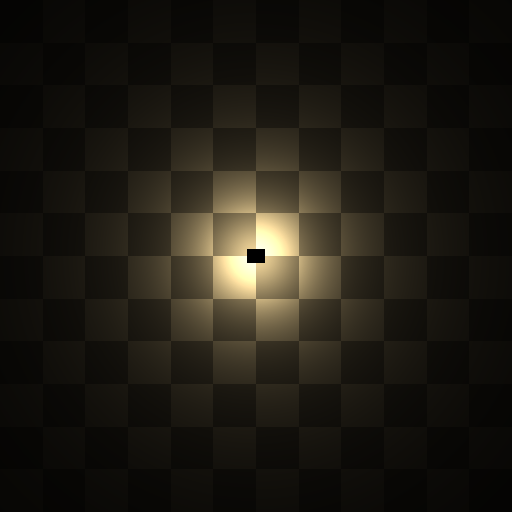} &
        \includegraphics[width=\linewidth]{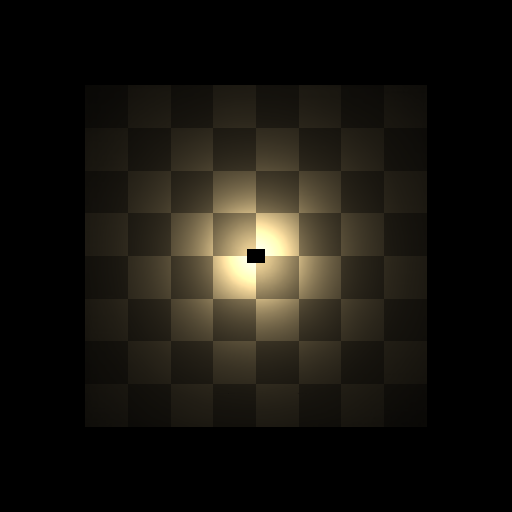} &
        \includegraphics[width=\linewidth]{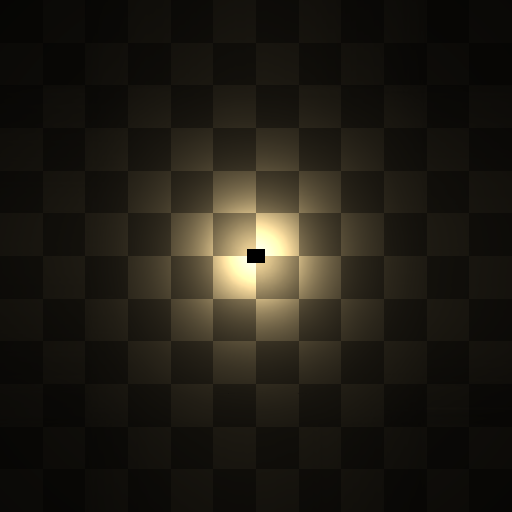} \\[6pt]
        \rotatebox{90}{\small Pierced sphere} &
        \includegraphics[width=\linewidth]{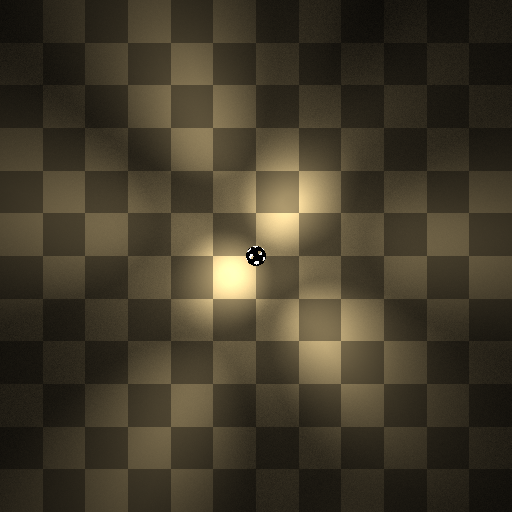} &
        \includegraphics[width=\linewidth]{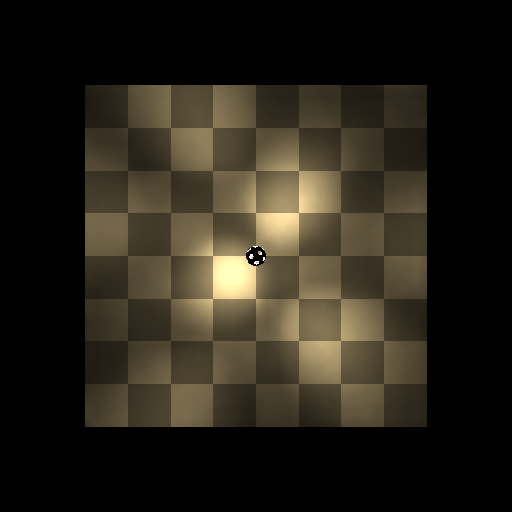} &
        \includegraphics[width=\linewidth]{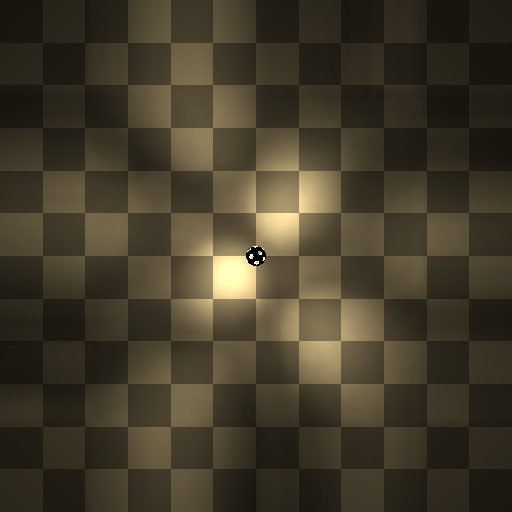} \\
    \end{tabular}
    \vspace{-1em}
    \caption{Emitters above a diffuse ground plane. The NEF trained in a bounded domain (center) drops to zero outside the AABB boundary. Applying an analytical fall-off from the boundary value (right) extends the field and approximates the reference (left) on areas outside of the NEF training domain.}
    \label{fig:pierced_plane}
\end{figure}

\textbf{Analytical Fall-off.}
A simple heuristic to deal with points outside of the emitter-space, consists of clamping outside queries to the nearest boundary point and scaling the resulting shading value by $(d_\text{boundary} / d_\text{actual})^2$, where $d_\text{boundary}$ is the distance from the emitter center to the clamped boundary point and $d_\text{actual}$ is the true distance from the emitter center to the shading point. As shown in \figref{fig:pierced_plane}, this fall-off approach generates a result that is hard to visually distinguish from the ground truth reference. Alternatively, the emitter-space can be enlarged at training time to avoid the heuristic, potentially requiring a larger representation to get the desired level of accuracy everywhere in the emitter-space. 

\begin{figure}
    \centering
    \includegraphics[width=\linewidth]{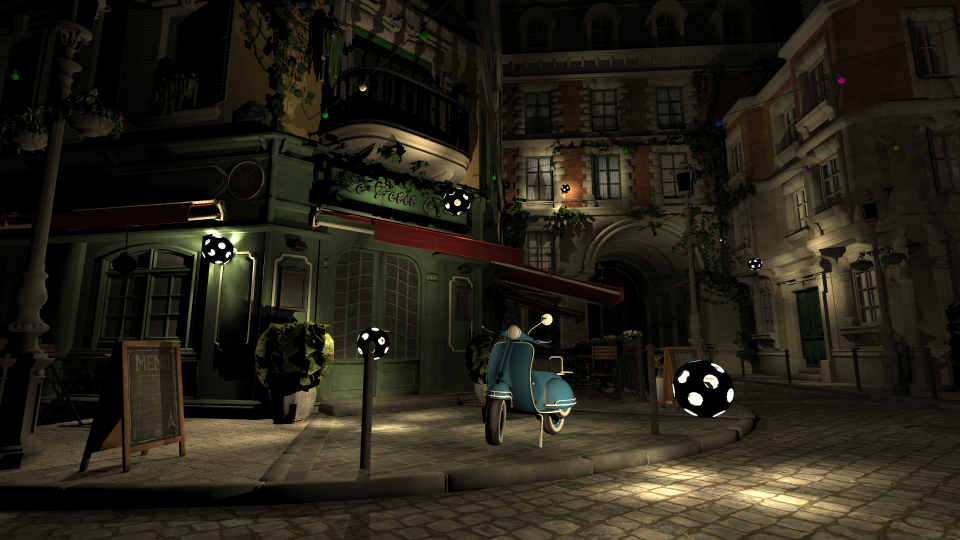}
    \vspace{-1em}
    \caption{Portability across instances. Six instances of the same \textit{Pierced Sphere} in the Bistro scene at different positions and orientations. A single trained NEF is loaded once; at each shading point, the world-space configuration $(x, n, \omega_o)$ is transformed into the emitter's local frame via the per-instance $M^{-1}$. Figure~\ref{fig:full_page_2} contains a comparison to reference for this scene.}
    \label{fig:bistro-instances}
\end{figure}

\textbf{Scene-independent placement.}
Anchoring the representation to the emitter's local frame yields two practical benefits: scene-independence and free rigid placement at runtime.
The trained NEF does not store scene-specific information, so the same parameters $\theta$ are reusable across scenes. At runtime, world-space shading configurations are mapped through $M^{-1}$ before being evaluated by $D_\theta$. Since $M$ is applied at shading time rather than baked into the network weights, translating or rotating the emitter is essentially free: it only changes the transformation applied to each query, not the representation itself. Uniform scaling is handled by folding the scale factor into $M$ and scaling the emitter-space extent accordingly; non-uniform scaling and shears are not supported, as they would distort the learned angular distribution. A single NEF therefore defines a family of emitters related by similarity transformations, amortizing the training cost across all scenes and instances in which the emitter is placed (\figref{fig:bistro-instances}).

\subsection{Diffuse--Glossy Separation} \label{sec:diffuse_glossy_sep}
Similarly to related work~\cite{rainer2022neural}, we decompose our neural representation into separate diffuse $f^\theta_d(\tilde{x},\tilde{n})$ and glossy $f_g^\theta(\tilde{x},\tilde{n},r(\tilde{\omega}_o, \tilde{n}),\alpha)$ components. We compute the total pre-integrated direct illumination $D_\theta$ at a given shading point $(\tilde{x}, \tilde{n}, \tilde{\omega}_o,\alpha)$ as their sum:
\begin{equation*}    D_\theta(\tilde{x},\tilde{n},\tilde{\omega}_o,\alpha) = \underbrace{\rho_d(x) \cdot f^\theta_d(\tilde{x},\tilde{n})}_{\text{diffuse}} + \underbrace{\rho_g(x)\cdot f_g^\theta(\tilde{x},\tilde{n},r(\tilde{\omega}_o, \tilde{n}),\alpha)}_{\text{glossy}}.
\end{equation*}
Here, tilde denotes properties in the local emitter-space defined in Section~\ref{sec:emitter_space}, $\rho_d$ and $\rho_g$ are factored-out diffuse and specular textures of the shading point $x$ which retain high-frequency texture details~\cite{muller2021real}, $\theta$ are the trainable parameters of the neural representation, $r(\tilde{\omega}_o, \tilde{n})$ is the reflected outgoing direction vector $\tilde{\omega}_o$ around the normal $\tilde{n}$, and $\alpha$ is the GGX roughness parameter. For emitters that are only expected to see diffuse surfaces at runtime, the glossy head $f_g^\theta$ can be omitted during training (and during shading).

All network inputs are normalized to $[0,1]^3$, and a learned multi-resolution hash encoding ~\cite{muller2022instant} is applied to both the position $\tilde{x}$ and the reflected direction $r(\tilde{\omega}_o, \tilde{n})$. 
Encoding the reflected direction is critical for capturing the sharp highlights and complex reflections characteristic of low-roughness receiver geometry (see supplemental material), consistent with findings in the field of novel view synthesis~\cite{verbin2022ref}. 
Both the glossy and diffuse head use a 4-layer MLP of width 128 with ReLU activations. More details on the network architecture can be found in the supplemental material.

\subsection{Training} \label{sec:train}

A schematic of the training procedure is shown in Figure~\ref{fig:training_and_forward}.
Training a NEF only requires a reliable estimate of the outgoing radiance
$L_o$ for any query within the network's input domain, making the procedure
agnostic to the underlying emitter geometry. 

\textbf{Sample generation.}
For each training batch, we draw queries uniformly from the emitter-space (\secref{sec:emitter_space}). For the glossy head
$f_g^\theta$, we sample a position $\tilde{x}$ uniformly in the local
bounding box, and a normal $\tilde{n}$ such that the emitter is visible in its hemisphere avoiding queries with zero radiance, an outgoing
direction $\tilde{\omega}_o$ uniformly on the upper hemisphere defined by
$\tilde{n}$, and a roughness $\alpha$ uniformly in $[0.1, 1]$. The diffuse
head $f_d^\theta$ uses the same procedure restricted to $(\tilde{x},
\tilde{n})$. For deformable or animated emitters, we additionally sample a
time parameter $t \in [0, 1]$ corresponding to the emitter's state, encoded
with an 8-band Fourier encoding~\cite{mildenhall2021nerf} and provided as
an extra conditioning input.

\textbf{Radiance targets.} 
Target radiance is obtained via offline path tracing
against the (possibly time-varying) emitter geometry. Bare emissive
surfaces are handled with standard next-event estimation; emitters with
non-emissive housing geometry that causes self-occlusion or internal
interreflections (\figref{fig:emissive_meshes}) are handled with full
path tracing inside the emitter assembly, so that higher-order transport
is naturally absorbed into the training signal. We use 64 samples per
radiance estimate as default across all emitters in this paper as we saw diminishing returns beyond this (see supplemental material for an ablation).
Targets are generated on-the-fly rather than cached, so all
reported training timings include the cost of path tracing.

\textbf{Optimization.}
We minimize a relative $L_2$ loss~\cite{muller2021real} between the network prediction and the path-traced target averaged over the batch,
\(
\mathcal{L}(\theta) =  \frac{(D_\theta - \langle L_o
\rangle)^2}{\mathrm{sg}(D_\theta)^2 + \epsilon} ,
\) where $\epsilon = 0.01$ and sg denotes a stop gradient operation.
The relative loss prevents bright regions of the emitter-space from dominating the
gradient and helps the multi-resolution hash encoding to
allocate capacity uniformly across the field. We use Adam~\cite{kingma2014adam}
with a learning rate of $10^{-3}$, a batch size of $2^{16}$, and use cosine annealing learning rate decay. The number of training iterations  for each of our emitter types can be found in the supplemental material.  Total wall-clock time on an RTX 4090
ranges from one minute for diffuse receiver with planar emitters to three hours
for emitters with complex emissive effects, dominated by the cost of
generating path-traced target values.

\subsection{Rendering} \label{sec:render}
Because NEF stores a pre-integrated estimate of the outgoing radiance reflected by the receiver, no integration is required at runtime. Unlike methods that integrate with respect to the emitter geometry or to a proxy representation of the emitter~\cite{heidrich1998canned, velazquez2015complex, zhu2021neural}, our approach reduces the shading process to a single network evaluation per pixel.
At render time, we first perform a G-buffer pass to obtain the $(x, n, \omega_o, \alpha)$ channels at each shading point. These are transformed into the emitter's local frame via its current world transform $M$ (\secref{sec:emitter_space}), encoded, and fed to $D_\theta$ to obtain the outgoing radiance contribution pre-integrated with respect to the emitter. 

\begin{figure}
    \centering
    \begin{minipage}[t]{0.329\linewidth}
        \centering
        \includegraphics[width=\linewidth]{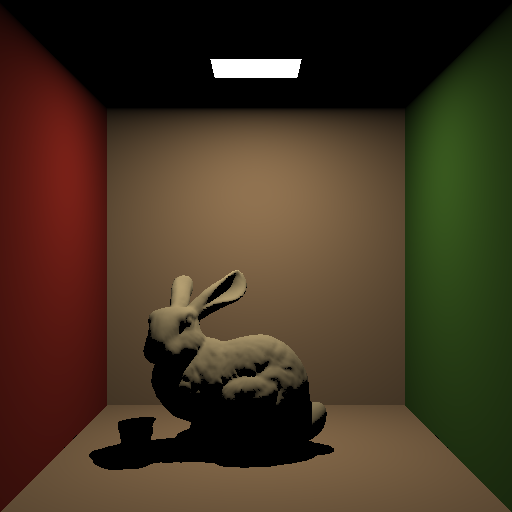}\\[2pt]
        {\small Reference (direct)}
    \end{minipage}\hfill
    \begin{minipage}[t]{0.329\linewidth}
        \centering
        \includegraphics[width=\linewidth]{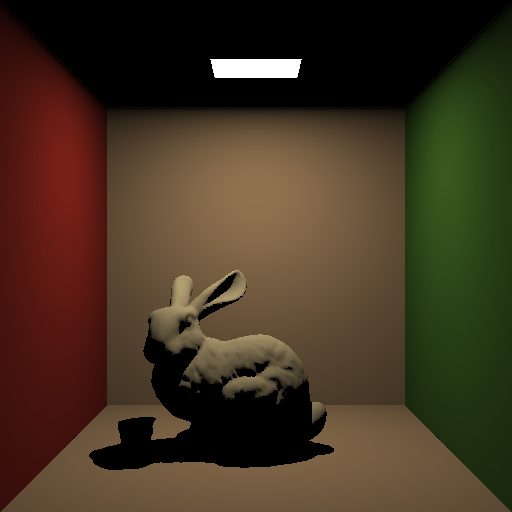}\\[2pt]
        {\small Analytical}
    \end{minipage}\hfill
    \begin{minipage}[t]{0.329\linewidth}
        \centering
        \includegraphics[width=\linewidth]{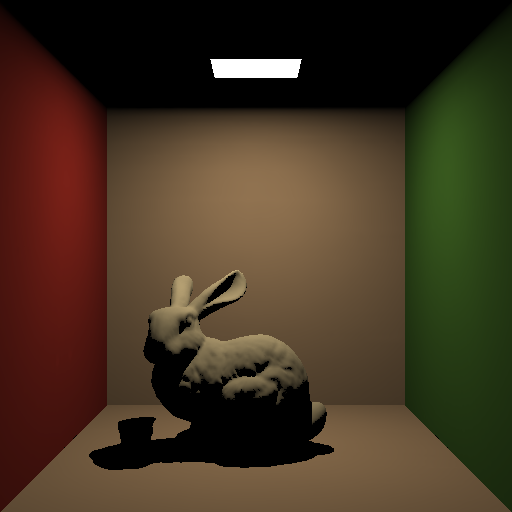}\\[2pt]
        {\small Ours (NEF) \\ \FLIP: 0.0041}
    \end{minipage}
    \caption{Comparison to analytical integration~\cite{baum1989} for a uniform planar emitter and diffuse receiver. Our method attains comparable accuracy without assuming any specific emitter geometry. Throughout the paper a hard shadow relative to the emitter center and the scene geometry is applied and \FLIP~\cite{Andersson2020} is computed on the LDR renders.}
    \label{fig:planar-uniform}
\end{figure}

\textbf{Visibility.} As with existing analytical methods~\cite{ltc2016, baum1989}, NEF does not inherently capture occlusions of the emissive field caused by the surrounding scene. For simple emitters, a single shadow ray toward the emitter center is often sufficient (\figref{fig:planar-uniform}); for soft shadows, established techniques such as shadow mapping~\cite{williams1978casting} or a small ray-tracing budget combined with a denoiser~\cite{heitz2018combining} apply directly, which we implemented in the supplemental material. Crucially, shadow rays are only required against non-emitter scene geometry: the NEF itself bakes in self-occlusion and internal interreflections \emph{within} the emitter assembly (\figref{fig:emissive_meshes}).
\section{Results}

We show that NEF matches the quality of analytical solutions (\secref{sec:planar}) and extends beyond what they handle, supporting emissive meshes and deformable emitters (\secref{sec:meshes}). We then compare against two sampling-based techniques, where NEF's sampling-free formulation yields a runtime advantage (\secref{sec:restir}).

\subsection{Planar Emitters} \label{sec:planar}
We compare against analytical integration for diffuse receivers~\cite{baum1989} and Linearly Transformed Cosines (LTC) for glossy receivers~\cite{ltc2016}.
As shown in \figref{fig:planar-uniform}, our method matches the analytical reference for diffuse receivers under uniform emission,  confirming its accuracy. 

\begin{figure*}
    \centering
    \setlength{\tabcolsep}{2pt}
    \begin{tabular}{@{}>{\centering\arraybackslash}m{0.04\linewidth} @{\hspace{4pt}} >{\centering\arraybackslash}m{0.185\linewidth} >{\centering\arraybackslash}m{0.185\linewidth} >{\centering\arraybackslash}m{0.185\linewidth} >{\centering\arraybackslash}m{0.185\linewidth} >{\centering\arraybackslash}m{0.185\linewidth}@{}}
        & {\small Path traced} & {\small LTC~\cite{ltc2016}} & {\small LTC error} & {\small NEF} & {\small NEF error} \\[2pt]
        \shortstack{$\alpha$\\[1pt]$0.10$} & \parbox[b]{\linewidth}{\centering\includegraphics[width=\linewidth]{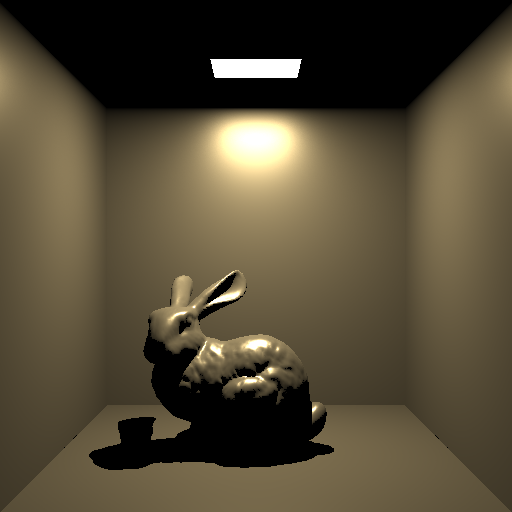}\\[1pt]{\footnotesize \shortstack{\phantom{\FLIP\ ($\downarrow$): 0.0000}\\\phantom{rMSE ($\downarrow$): 0.0000}}}} & \parbox[b]{\linewidth}{\centering\includegraphics[width=\linewidth]{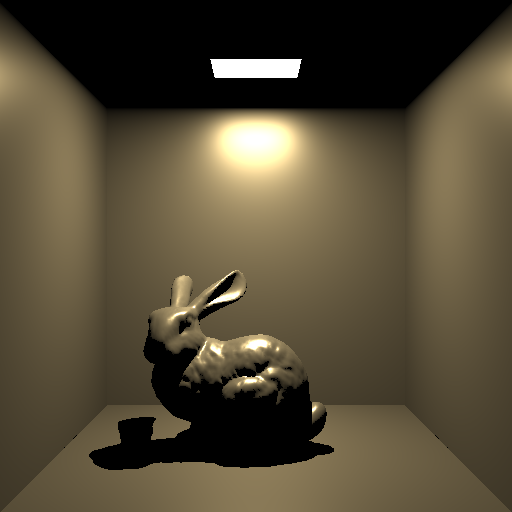}\\[1pt]{\footnotesize \shortstack{\FLIP\ ($\downarrow$): 0.0459\\rMSE ($\downarrow$): 0.0013}}} & \parbox[b]{\linewidth}{\centering\includegraphics[width=\linewidth]{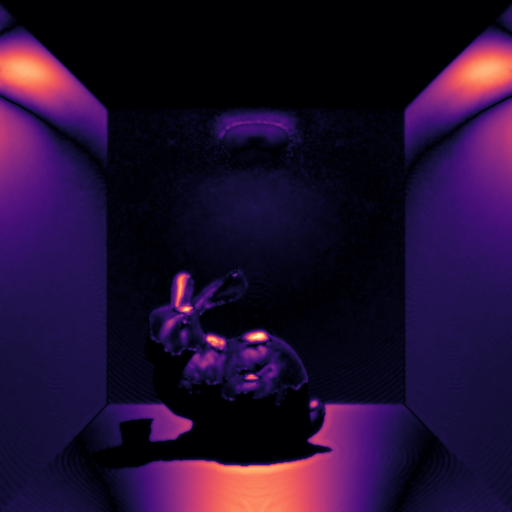}\\[1pt]{\footnotesize \shortstack{\phantom{\FLIP\ ($\downarrow$): 0.0000}\\\phantom{rMSE ($\downarrow$): 0.0000}}}} & \parbox[b]{\linewidth}{\centering\includegraphics[width=\linewidth]{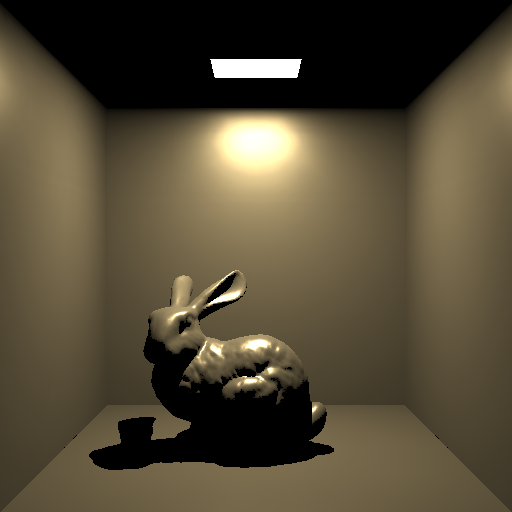}\\[1pt]{\footnotesize \shortstack{\FLIP\ ($\downarrow$): \textbf{0.0188}\\rMSE ($\downarrow$): \textbf{0.0001}}}} & \parbox[b]{\linewidth}{\centering\includegraphics[width=\linewidth]{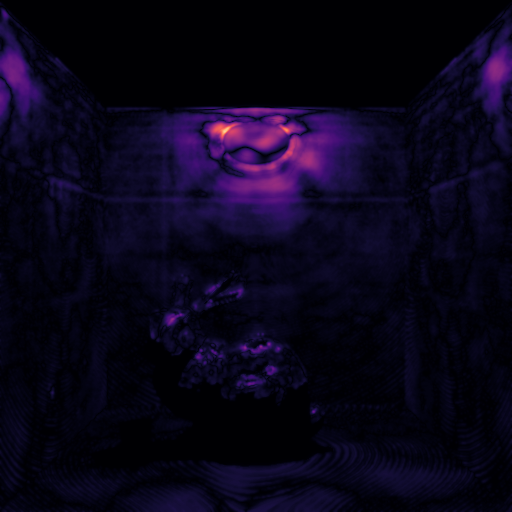}\\[1pt]{\footnotesize \shortstack{\phantom{\FLIP\ ($\downarrow$): 0.0000}\\\phantom{rMSE ($\downarrow$): 0.0000}}}} \\[14pt]
        \shortstack{$\alpha$\\[1pt]$0.30$} & \parbox[b]{\linewidth}{\centering\includegraphics[width=\linewidth]{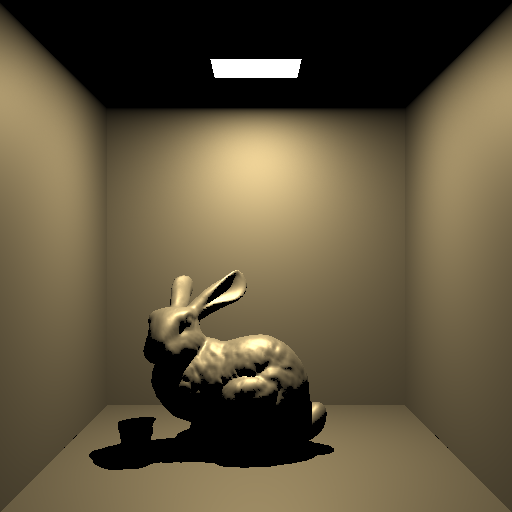}\\[1pt]{\footnotesize \shortstack{\phantom{\FLIP\ ($\downarrow$): 0.0000}\\\phantom{rMSE ($\downarrow$): 0.0000}}}} & \parbox[b]{\linewidth}{\centering\includegraphics[width=\linewidth]{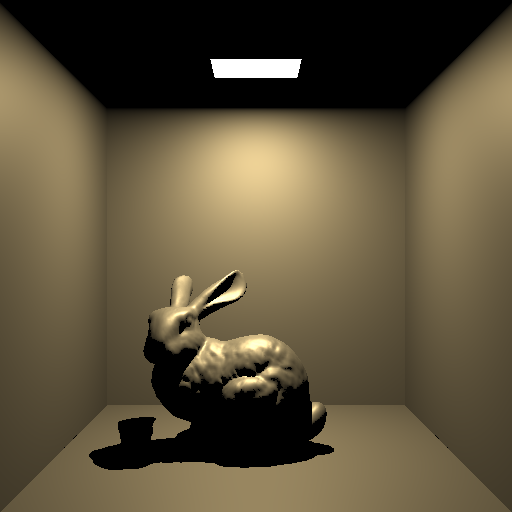}\\[1pt]{\footnotesize \shortstack{\FLIP\ ($\downarrow$): 0.0716\\rMSE ($\downarrow$): 0.0014}}} & \parbox[b]{\linewidth}{\centering\includegraphics[width=\linewidth]{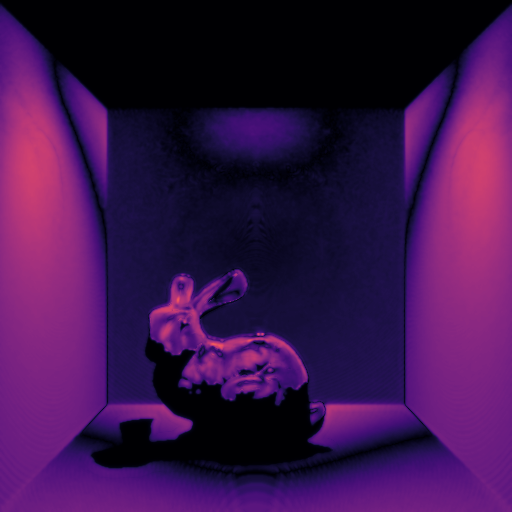}\\[1pt]{\footnotesize \shortstack{\phantom{\FLIP\ ($\downarrow$): 0.0000}\\\phantom{rMSE ($\downarrow$): 0.0000}}}} & \parbox[b]{\linewidth}{\centering\includegraphics[width=\linewidth]{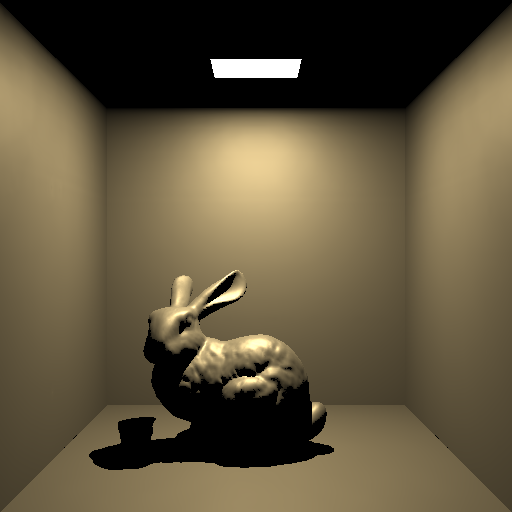}\\[1pt]{\footnotesize \shortstack{\FLIP\ ($\downarrow$): \textbf{0.0169}\\rMSE ($\downarrow$): \textbf{0.0001}}}} & \parbox[b]{\linewidth}{\centering\includegraphics[width=\linewidth]{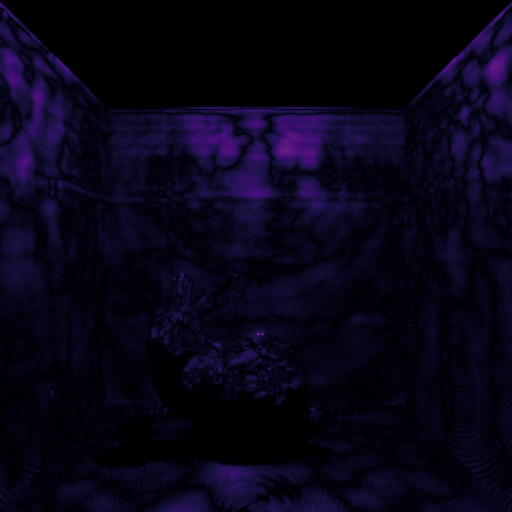}\\[1pt]{\footnotesize \shortstack{\phantom{\FLIP\ ($\downarrow$): 0.0000}\\\phantom{rMSE ($\downarrow$): 0.0000}}}} \\[14pt]
        \shortstack{$\alpha$\\[1pt]$0.50$} & \parbox[b]{\linewidth}{\centering\includegraphics[width=\linewidth]{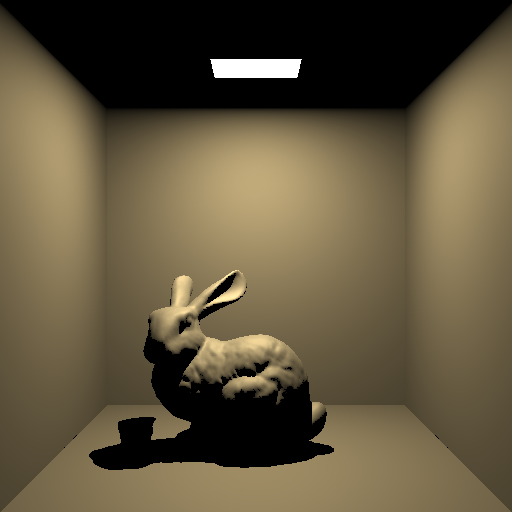}\\[1pt]{\footnotesize \shortstack{\phantom{\FLIP\ ($\downarrow$): 0.0000}\\\phantom{rMSE ($\downarrow$): 0.0000}}}} & \parbox[b]{\linewidth}{\centering\includegraphics[width=\linewidth]{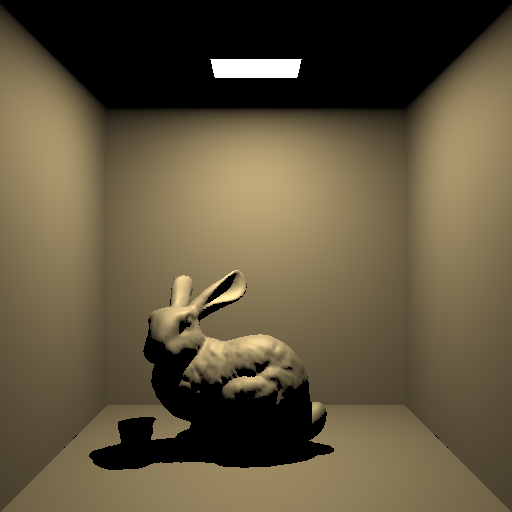}\\[1pt]{\footnotesize \shortstack{\FLIP\ ($\downarrow$): 0.0470\\rMSE ($\downarrow$): 0.0004}}} & \parbox[b]{\linewidth}{\centering\includegraphics[width=\linewidth]{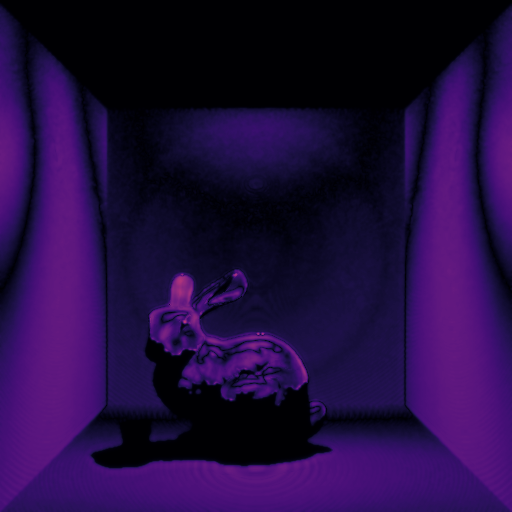}\\[1pt]{\footnotesize \shortstack{\phantom{\FLIP\ ($\downarrow$): 0.0000}\\\phantom{rMSE ($\downarrow$): 0.0000}}}} & \parbox[b]{\linewidth}{\centering\includegraphics[width=\linewidth]{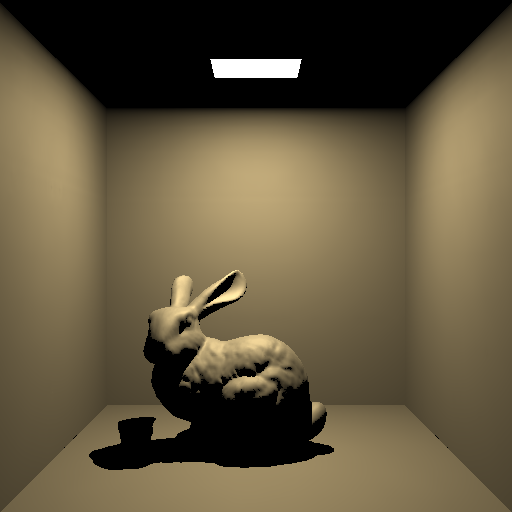}\\[1pt]{\footnotesize \shortstack{\FLIP\ ($\downarrow$): \textbf{0.0149}\\rMSE ($\downarrow$): \textbf{0.0001}}}} & \parbox[b]{\linewidth}{\centering\includegraphics[width=\linewidth]{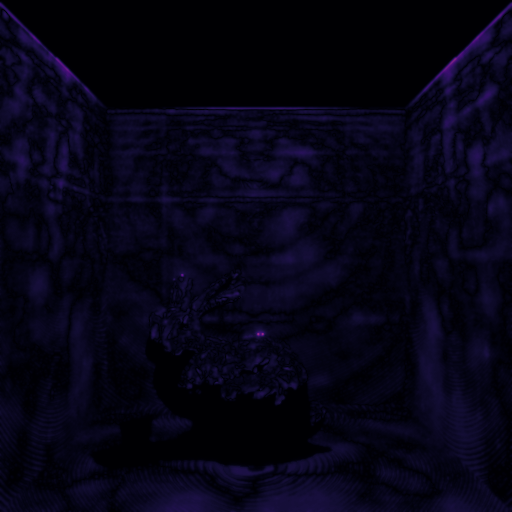}\\[1pt]{\footnotesize \shortstack{\phantom{\FLIP\ ($\downarrow$): 0.0000}\\\phantom{rMSE ($\downarrow$): 0.0000}}}} \\[14pt]
        \shortstack{$\alpha$\\[1pt]$0.70$} & \parbox[b]{\linewidth}{\centering\includegraphics[width=\linewidth]{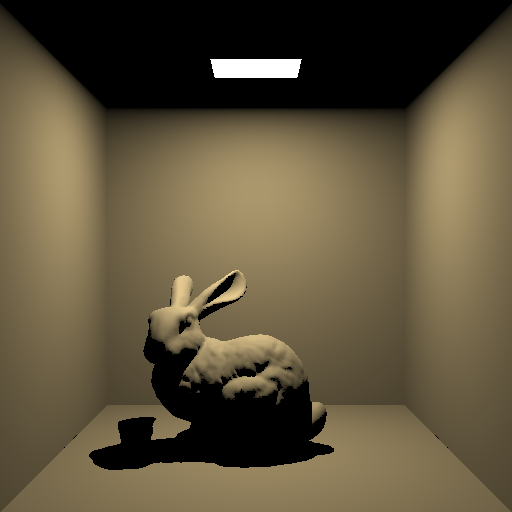}\\[1pt]{\footnotesize \shortstack{\phantom{\FLIP\ ($\downarrow$): 0.0000}\\\phantom{rMSE ($\downarrow$): 0.0000}}}} & \parbox[b]{\linewidth}{\centering\includegraphics[width=\linewidth]{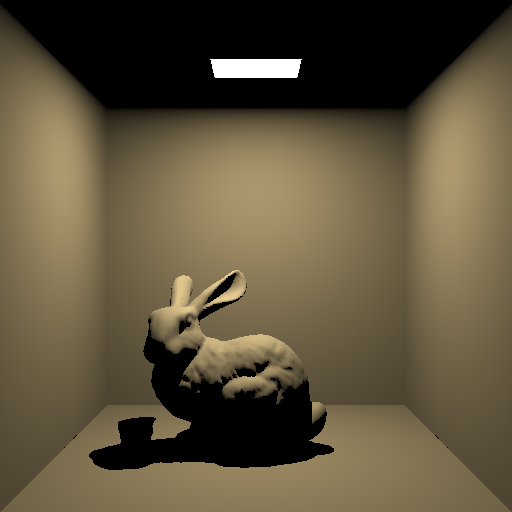}\\[1pt]{\footnotesize \shortstack{\FLIP\ ($\downarrow$): 0.0271\\rMSE ($\downarrow$): 0.0001}}} & \parbox[b]{\linewidth}{\centering\includegraphics[width=\linewidth]{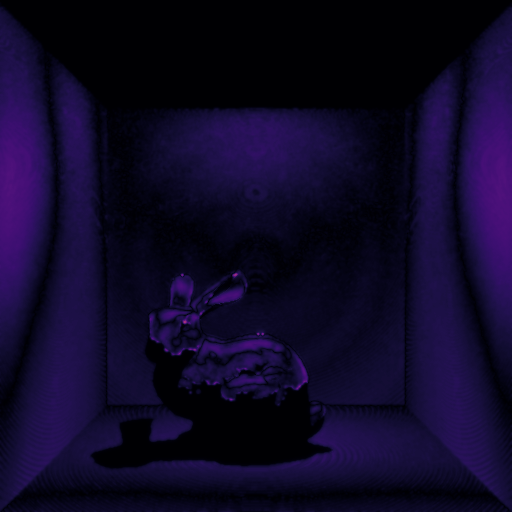}\\[1pt]{\footnotesize \shortstack{\phantom{\FLIP\ ($\downarrow$): 0.0000}\\\phantom{rMSE ($\downarrow$): 0.0000}}}} & \parbox[b]{\linewidth}{\centering\includegraphics[width=\linewidth]{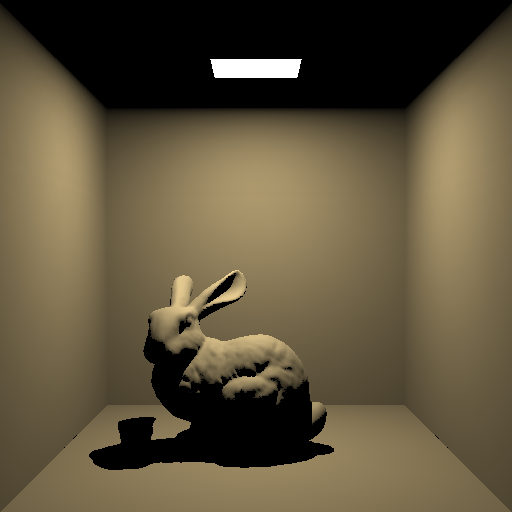}\\[1pt]{\footnotesize \shortstack{\FLIP\ ($\downarrow$): \textbf{0.0128}\\rMSE ($\downarrow$): \textbf{0.0001}}}} & \parbox[b]{\linewidth}{\centering\includegraphics[width=\linewidth]{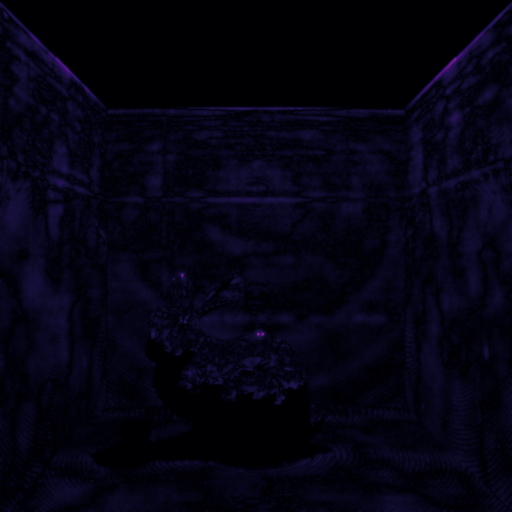}\\[1pt]{\footnotesize \shortstack{\phantom{\FLIP\ ($\downarrow$): 0.0000}\\\phantom{rMSE ($\downarrow$): 0.0000}}}} \\[14pt]
        \shortstack{$\alpha$\\[1pt]$0.90$} & \parbox[b]{\linewidth}{\centering\includegraphics[width=\linewidth]{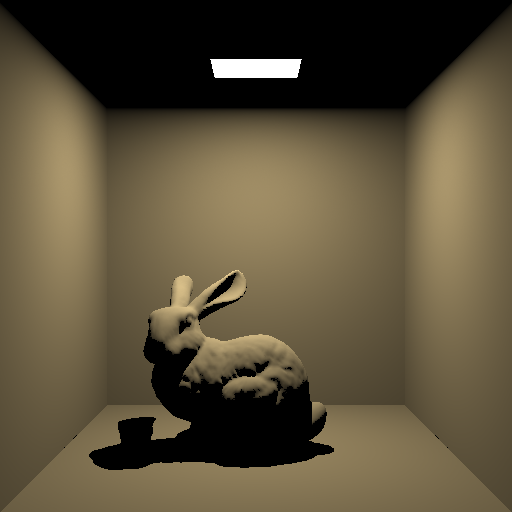}\\[1pt]{\footnotesize \shortstack{\phantom{\FLIP\ ($\downarrow$): 0.0000}\\\phantom{rMSE ($\downarrow$): 0.0000}}}} & \parbox[b]{\linewidth}{\centering\includegraphics[width=\linewidth]{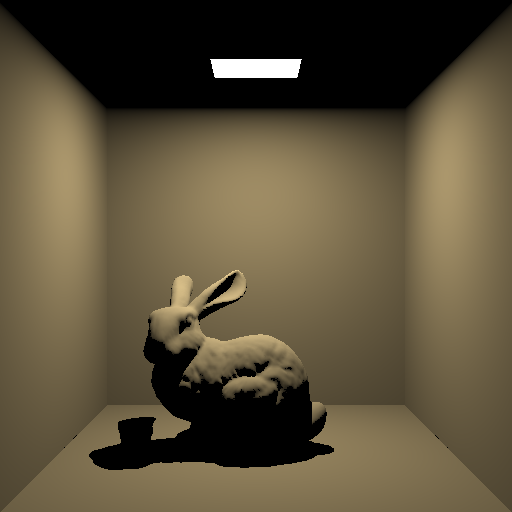}\\[1pt]{\footnotesize \shortstack{\FLIP\ ($\downarrow$): 0.0191\\rMSE ($\downarrow$): 0.0001}}} & \parbox[b]{\linewidth}{\centering\includegraphics[width=\linewidth]{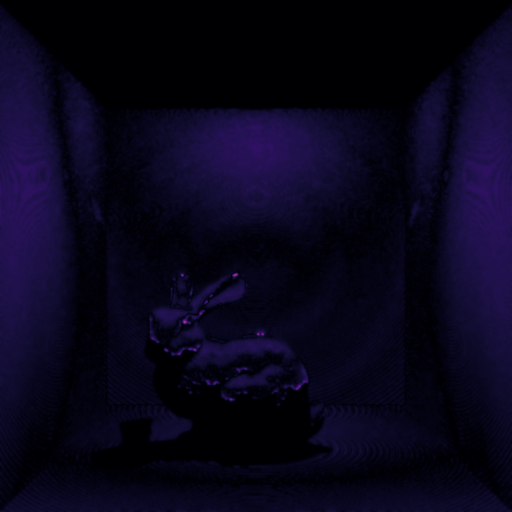}\\[1pt]{\footnotesize \shortstack{\phantom{\FLIP\ ($\downarrow$): 0.0000}\\\phantom{rMSE ($\downarrow$): 0.0000}}}} & \parbox[b]{\linewidth}{\centering\includegraphics[width=\linewidth]{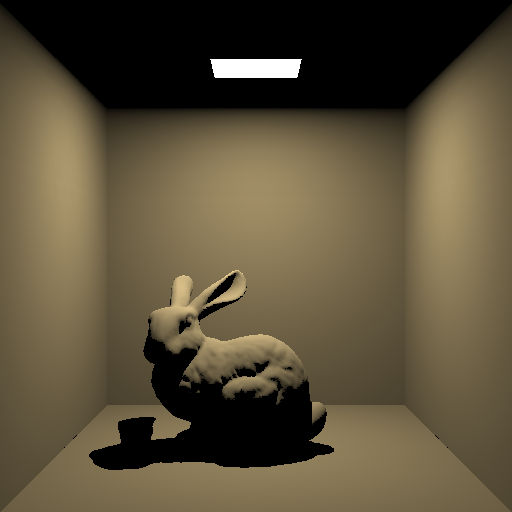}\\[1pt]{\footnotesize \shortstack{\FLIP\ ($\downarrow$): \textbf{0.0125}\\rMSE ($\downarrow$): \textbf{0.0001}}}} & \parbox[b]{\linewidth}{\centering\includegraphics[width=\linewidth]{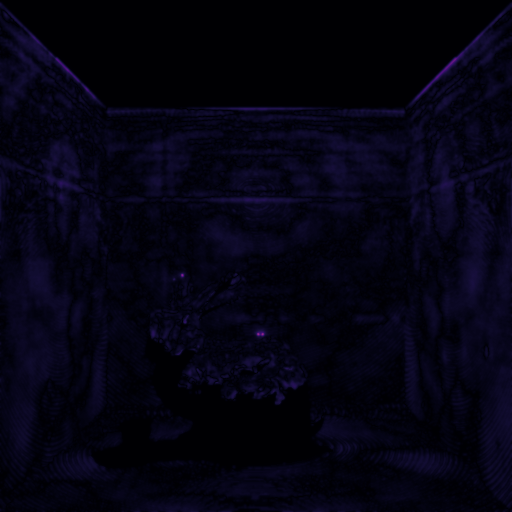}\\[1pt]{\footnotesize \shortstack{\phantom{\FLIP\ ($\downarrow$): 0.0000}\\\phantom{rMSE ($\downarrow$): 0.0000}}}}
        \\
    \end{tabular}
    \caption{Linearly Transformed Cosines~\cite{ltc2016} permits rendering the direct illumination of simple emitters like the polygonal light source shown above, for glossy receivers like a GGX specular material~\cite{walter2007microfacet}, where $\alpha$ is the roughness parameter. However, the method's approximate nature prevents a perfect match with a path-traced solution. By learning the pre-integrated emission around the emitter with our NEF, we obtain a better fit without losing generality across supported emitter types. All \FLIP{} maps share a common color scale.}
    \label{fig:ltc_vs_nef}
\end{figure*}

For glossy receivers (\figref{fig:ltc_vs_nef}), the larger capacity of NEF allows it to achieve lower rendering error across various roughness values $\alpha$ compared to LTC, highlighting LTC's remaining approximation error.
The near-mirror regime ($\alpha \to 0$) remains challenging for a finite neural representation; we show an example of the increase in error at lower roughness in the supplemental material.

While analytical methods have lower per-frame shading costs for simple polygons, their runtime scales linearly $\mathcal{O}(V)$ with the number of emitter vertices $V$. This linear cost is especially noticeable for large emissive emitters, as highlighted in \tabref{tab:runtime}. We show the crossover point between the analytical methods and our GlossyNEF---267 emitter vertices---in the supplemental material. The generality and computational efficiency of our approach become especially relevant when extending beyond planar geometry.

\begin{table}[t]
    \centering
    \small
        \caption{Isolated direct illumination runtime cost at $1920 \times 1080$. Analytical and LTC scale linearly with vertex count; NEF cost is independent of emitter complexity. Times reflect the shading step only.
        We employ a standardized NEF architecture across all emitters rather than tuning hyperparameters for individual cases; consequently, the reported timings are consistent and representative across all figures and ran on an RTX 4090.}
    \begin{tabular}{lccc}
        \toprule
        & & Quad & Dragon \\
        Method & Component & (4 verts) & (436k verts) \\
        \midrule
        Analytical~\cite{baum1989} & Diffuse & 0.03 ms & 1951.4 ms \\
        LTC~\cite{ltc2016} & Glossy & 0.07 ms & 3925.6 ms \\
        Ours (NEF) & Diffuse & 1.2 ms & 1.2 ms \\
        Ours (GlossyNEF) & Glossy & 2.5 ms & 2.5 ms \\
        \bottomrule
    \end{tabular}
    \label{tab:runtime}
\end{table}

\subsection{Complex Emitters} \label{sec:meshes}

A key advantage of our approach is that inference cost is determined by the complexity of the emissive field, not by the complexity of the underlying geometry. While NEF inference cost does scale with emissive field complexity---a highly spatially varying emitter benefits from a larger or more expressive network---this complexity is typically much lower than that of the geometry itself. Naively subdividing an emissive mesh increases its vertex count but leaves its emissive field largely unchanged, so the same NEF capacity suffices regardless of tessellation. All results in this paper share a single network configuration sized for the most demanding case (the \textit{Pierced Sphere}, whose housing interreflections produce a high-frequency emissive field); for simpler emitters such as the planar quad, a smaller architecture would suffice with no perceptible loss in quality (see supplemental material).

\begin{figure}
    \centering
    \setlength{\tabcolsep}{2pt}
    \begin{tabular}{@{}>{\centering\arraybackslash}m{0.06\linewidth} @{\hspace{4pt}} >{\centering\arraybackslash}m{0.29\linewidth} >{\centering\arraybackslash}m{0.29\linewidth} >{\centering\arraybackslash}m{0.29\linewidth}@{}}
        & {\small Dragon} & {\small Pierced Sphere} & {\small Lattice Box} \\[2pt]
        \rotatebox{90}{\small Standard DI}
        & \includegraphics[width=\linewidth]{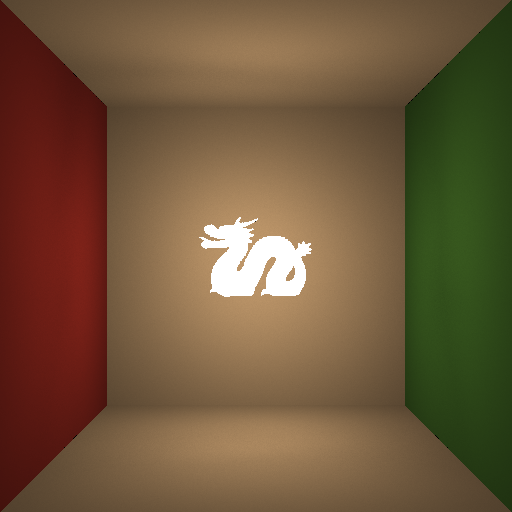}
        & \includegraphics[width=\linewidth]{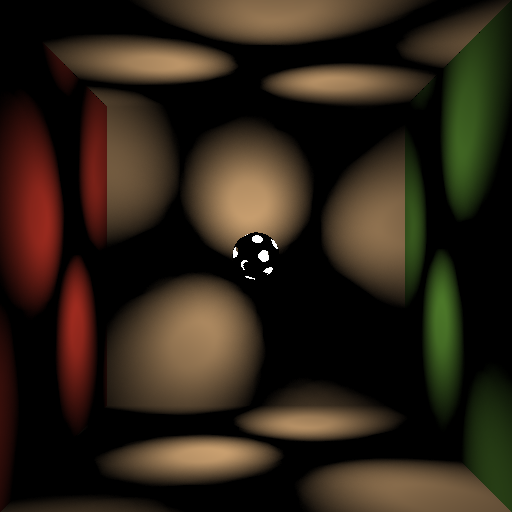}
        & \includegraphics[width=\linewidth]{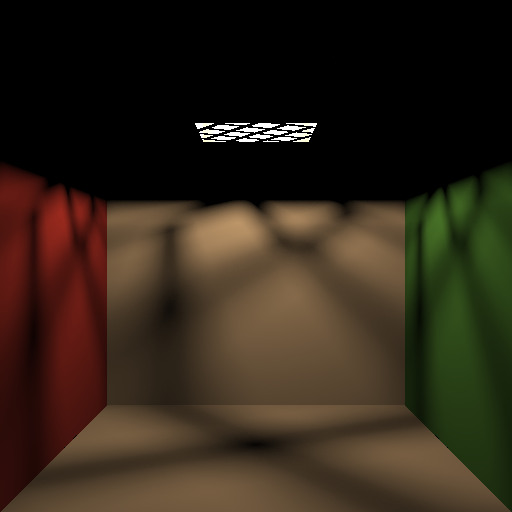} \\[4pt]
        \rotatebox{90}{\small + Interreflections}
        & \includegraphics[width=\linewidth]{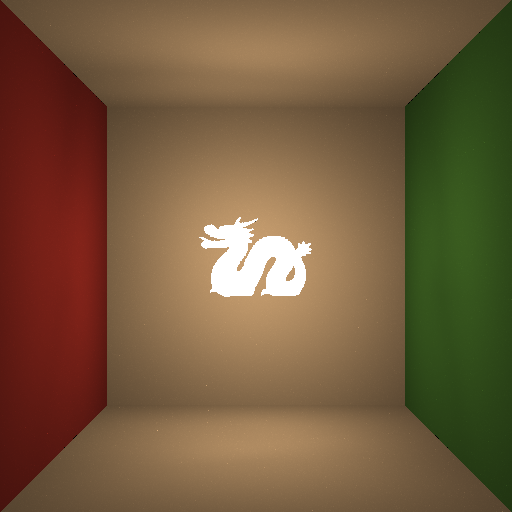}
        & \includegraphics[width=\linewidth]{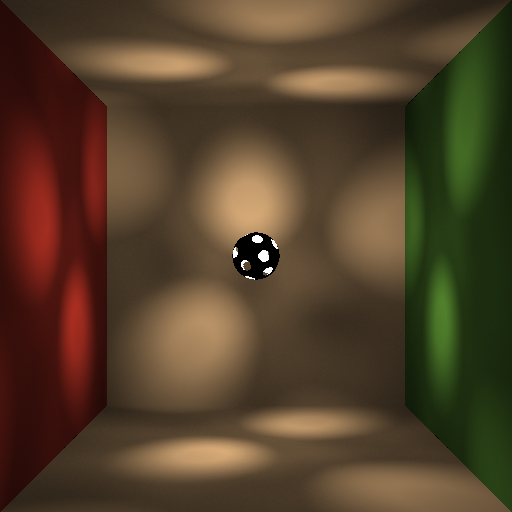}
        & \includegraphics[width=\linewidth]{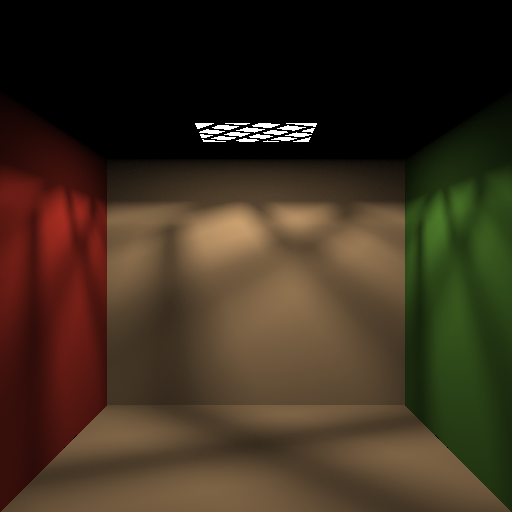} \\[4pt]
        \rotatebox{90}{\small NEF (ours)}
        & \includegraphics[width=\linewidth]{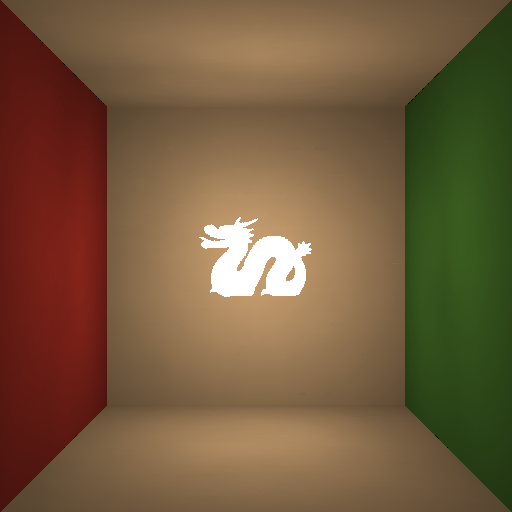}
        & \includegraphics[width=\linewidth]{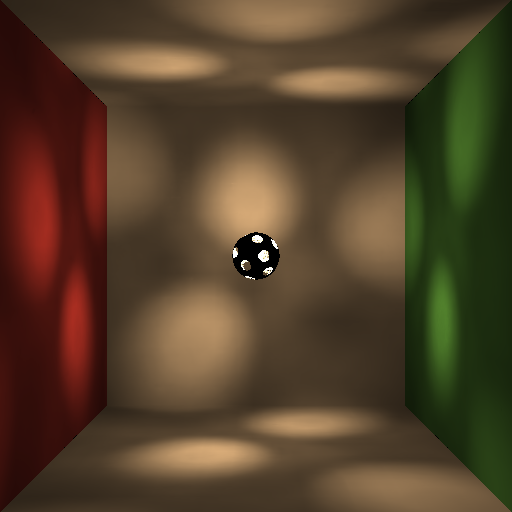}
        & \includegraphics[width=\linewidth]{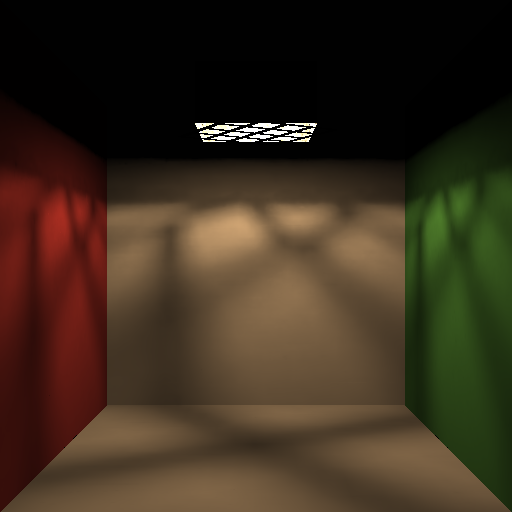} \\
    \end{tabular}
    \caption{Direct illumination from complex emitter assemblies. \emph{Top:} standard direct illumination without interreflections. \emph{Middle:} emitter-DI reference that includes bounces within the emitter assembly but excludes scene indirect. \emph{Bottom:} NEF predictions. }
    \label{fig:emissive_meshes}
\end{figure}

\textbf{Higher-order Emissive Effects.} 
Our NEF formulation can also capture higher-order emissive effects arising from multi-bounce light transport within the emitter, such as internal reflections toward a filament. By replacing direct light-source sampling with path tracing during training, multi-bounce contributions are integrated into the training signal.
\figref{fig:emissive_meshes} shows three assemblies in which internal interreflections meaningfully shape the emissive field: a \textit{Dragon} mesh acting as a bare emitter, a \textit{Pierced Sphere} whose shell partially occludes and reflects an internal filament, and a \textit{Lattice Box} which houses a spherical emitter. Obtaining a noise-free reference for these assemblies is expensive---on the order of $1\mathrm{M}$ samples per pixel---because the interior light paths are not captured by simple next-event estimation against the bare emissive geometry. 
The effect varies by assembly: interreflections produce only a mild overall brightening on the \textit{Dragon}, whereas the \textit{Pierced Sphere} and \textit{Lattice Box} show light diffusing through the apertures, softening cast shadows and noticeably brightening regions that the bare-emitter leaves dark. NEF reproduces all three references faithfully.

\textbf{Non-uniform emission.}
Since our NEF representation is agnostic to the emissive profile of the light source, it naturally accommodates spatially varying emission without modification. \figref{fig:teaser} (center) shows an emitter with a non-uniform radiance distribution.
While some prior methods support textured emitters, they rely on approximations that introduce error~\cite{ltc2016}.

\begin{figure}
    \centering
    \makebox[\linewidth][r]{%
        \makebox[0.94\linewidth][c]{%
            \begin{tikzpicture}
                \draw[->] (0,0) -- (0.94\linewidth,0) node[midway, above] {\small Time};
            \end{tikzpicture}%
        }%
    }

    \vspace{4pt}
    \setlength{\tabcolsep}{2pt}
    \begin{tabular}{@{}>{\centering\arraybackslash}m{0.04\linewidth} @{\hspace{4pt}} >{\centering\arraybackslash}m{0.30\linewidth} >{\centering\arraybackslash}m{0.30\linewidth} >{\centering\arraybackslash}m{0.30\linewidth}@{}}
        \rotatebox{90}{\small Reference} & \includegraphics[width=\linewidth]{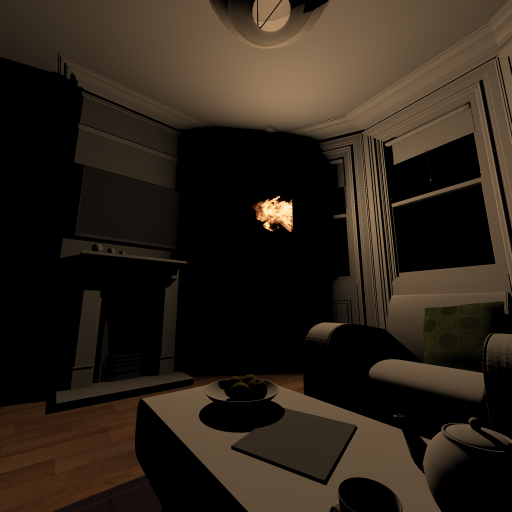} & \includegraphics[width=\linewidth]{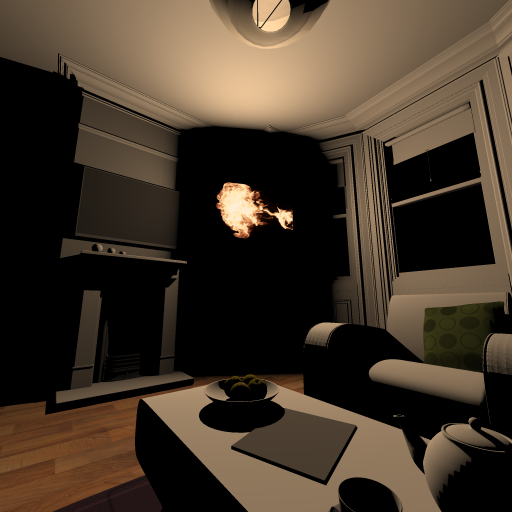} & \includegraphics[width=\linewidth]{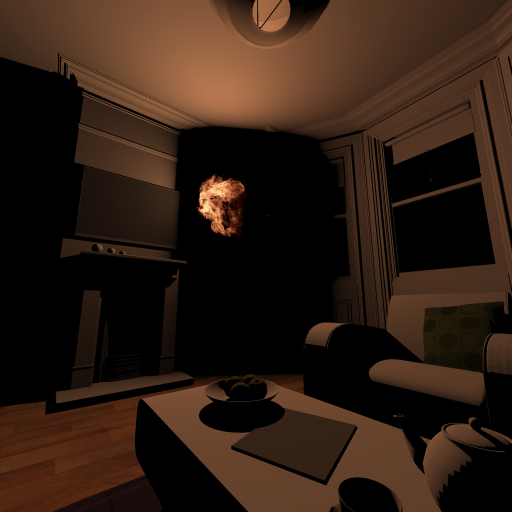} \\[6pt]
        \rotatebox{90}{\small NEF} & \includegraphics[width=\linewidth]{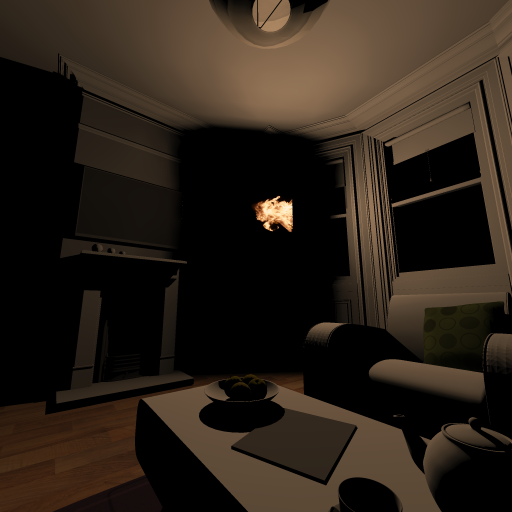} & \includegraphics[width=\linewidth]{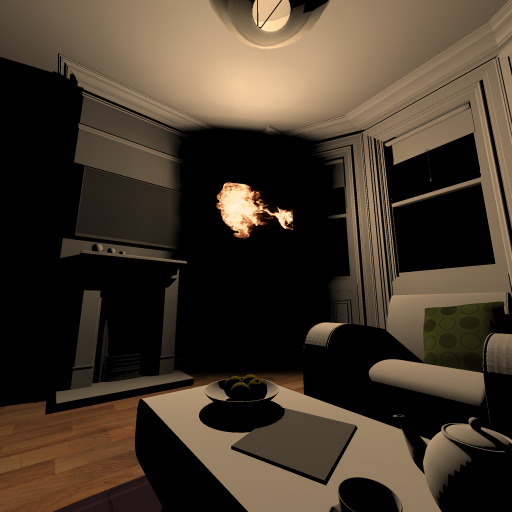} & \includegraphics[width=\linewidth]{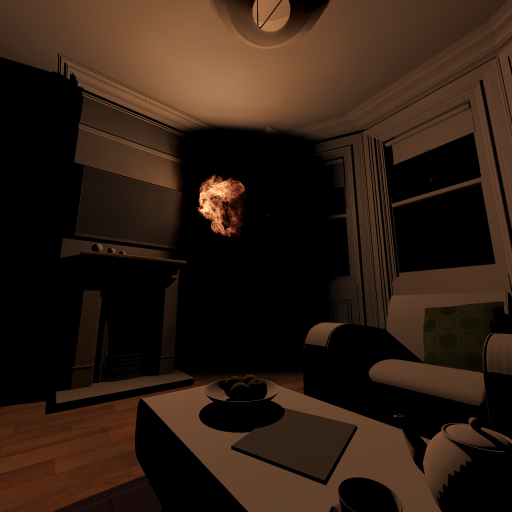} \\[2pt]
        & {\footnotesize \FLIP: 0.0525} & {\footnotesize \FLIP: 0.0599} & {\footnotesize \FLIP: 0.0358} \\
    \end{tabular}
    \caption{A textured, time-varying fireball emitter in the whiteroom scene at three frames of its animation, with the time parameter $t$ indexing into video frames. Top: path-traced reference. Bottom: NEF prediction from a single network evaluation per pixel. 
    }
    \label{fig:deformable-emitter}
\end{figure}

\begin{figure*}[t]
  \centering
  \setlength{\tabcolsep}{2pt}
  \begin{tabular}{@{}>{\centering\arraybackslash}m{0.02\linewidth} @{\hspace{4pt}} >{\centering\arraybackslash}m{0.31\linewidth} >{\centering\arraybackslash}m{0.31\linewidth} >{\centering\arraybackslash}m{0.31\linewidth}@{}}
    & {\small Reference} & {\small NEF} & {\small \FLIP} \\[4pt]
    \rotatebox{90}{\small Sponza}
    & \includegraphics[width=\linewidth]{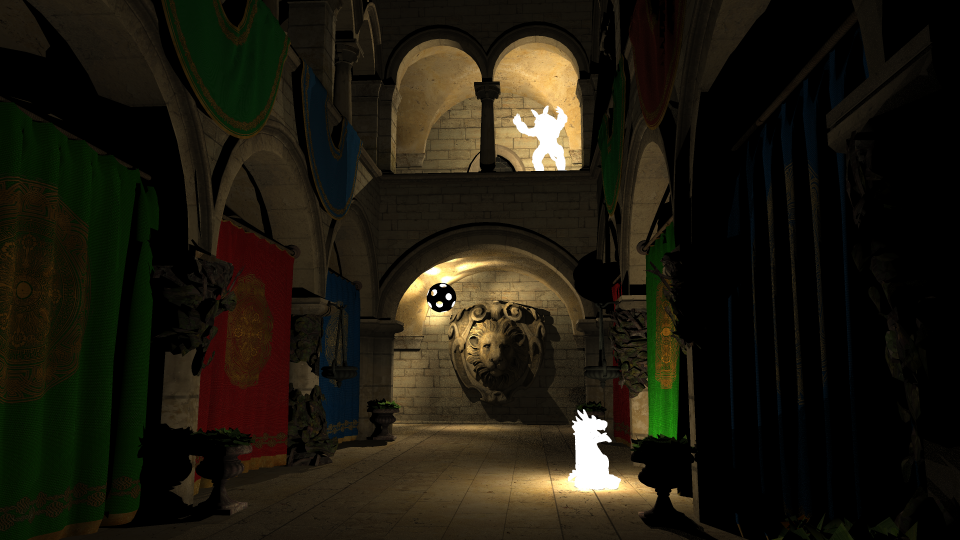}
    & \includegraphics[width=\linewidth]{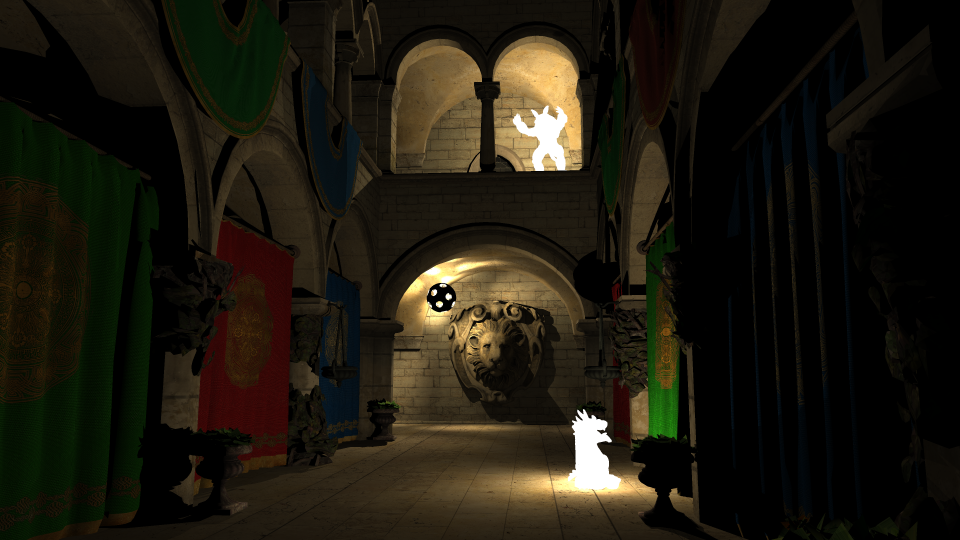}
    & \includegraphics[width=\linewidth]{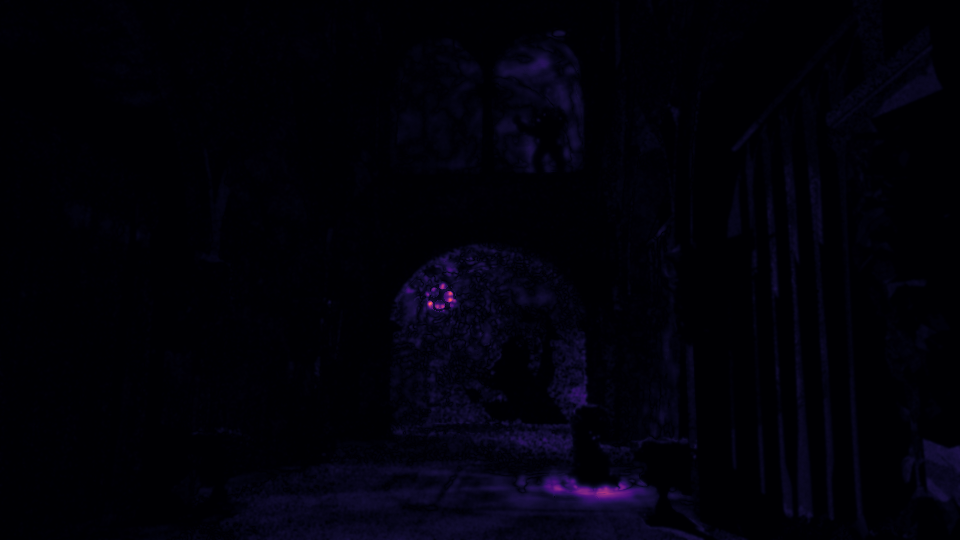} \\[-2pt]
    & & & {\normalsize \FLIP \ ($\downarrow$): 0.0137} \\[6pt]
    \rotatebox{90}{\small Bistro}
    & \includegraphics[width=\linewidth]{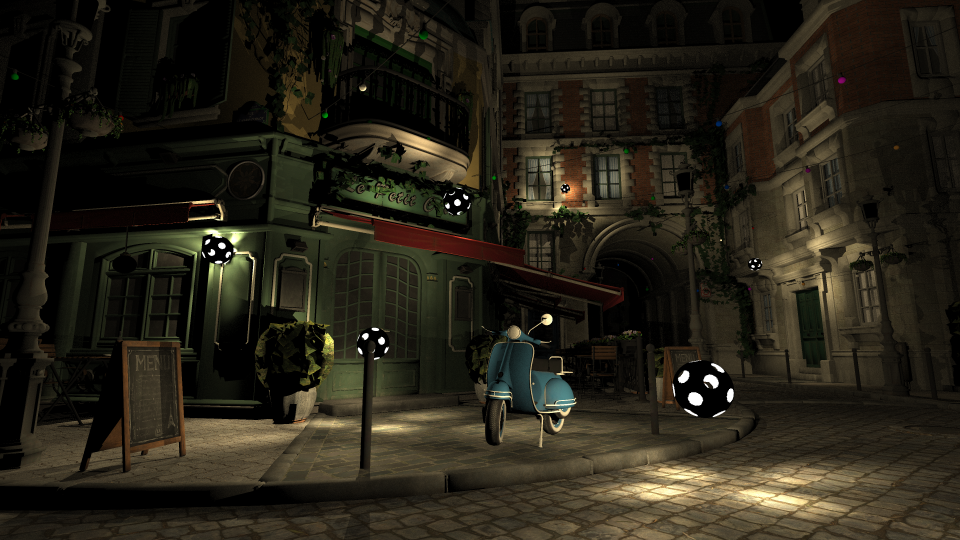}
    & \includegraphics[width=\linewidth]{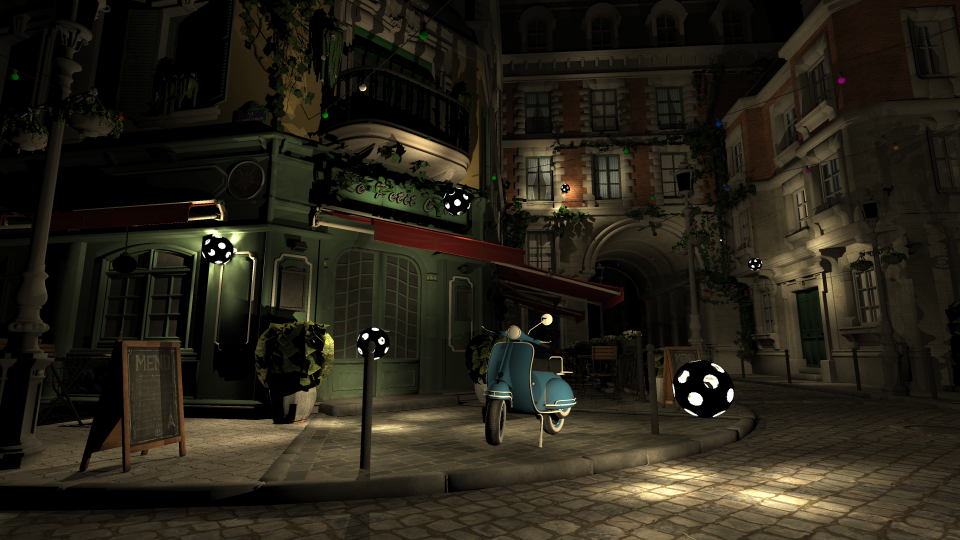}
    & \includegraphics[width=\linewidth]{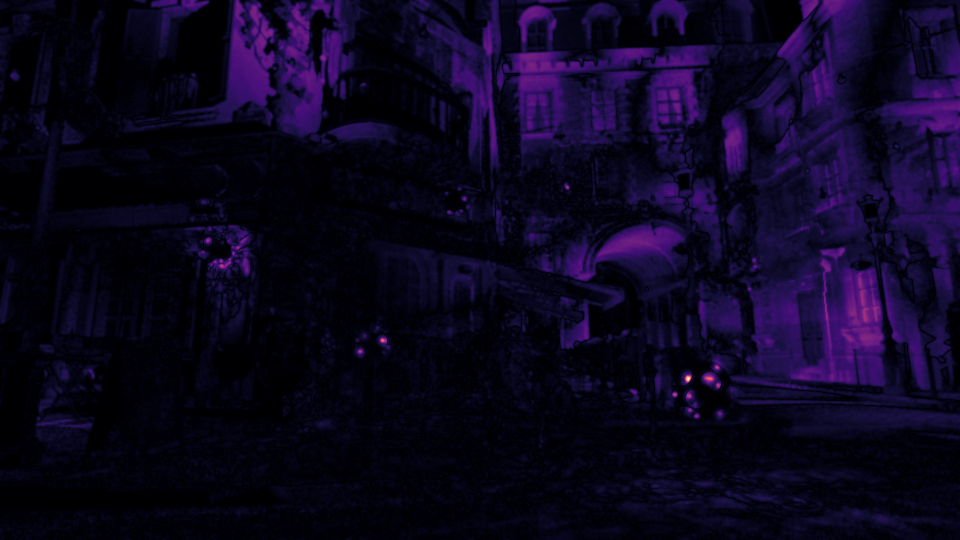} \\[-2pt]
    & & & {\normalsize \FLIP \ ($\downarrow$): 0.0536 } \\[6pt]
    \rotatebox{90}{\small Neon NEF}
    & \includegraphics[width=\linewidth]{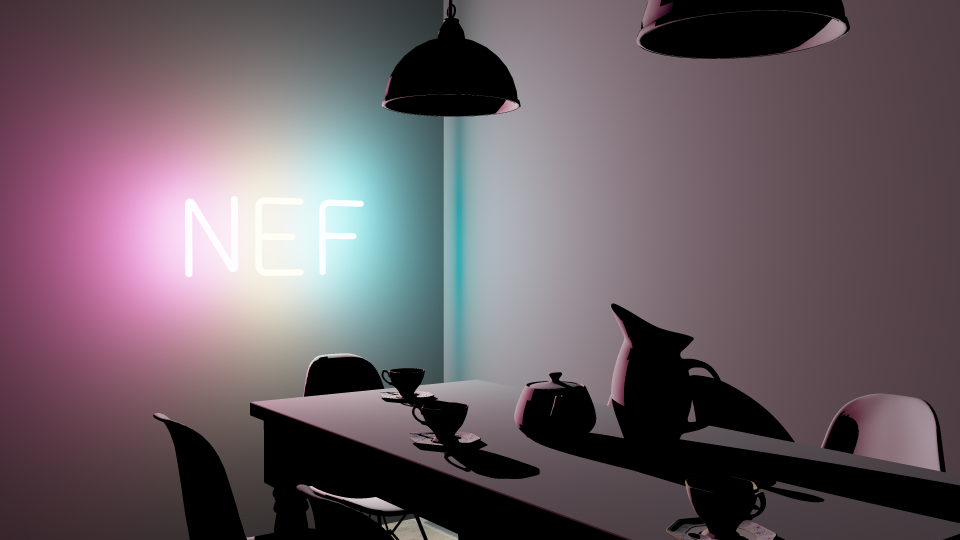}
    & \includegraphics[width=\linewidth]{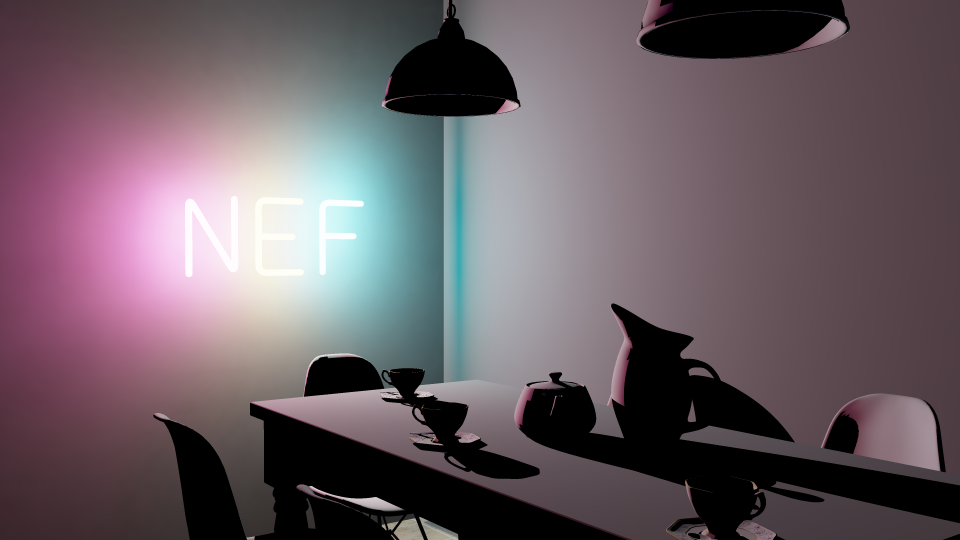}
    & \includegraphics[width=\linewidth]{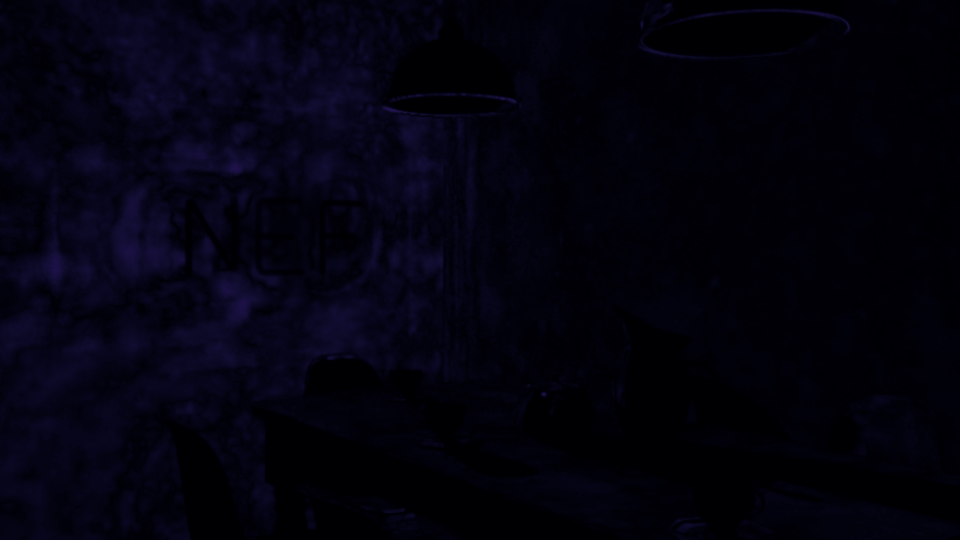} \\[-2pt]
    & & & {\normalsize \FLIP \ ($\downarrow$): 0.0252 } \\[6pt]
    \rotatebox{90}{\small Whiteroom}
    & \includegraphics[width=\linewidth]{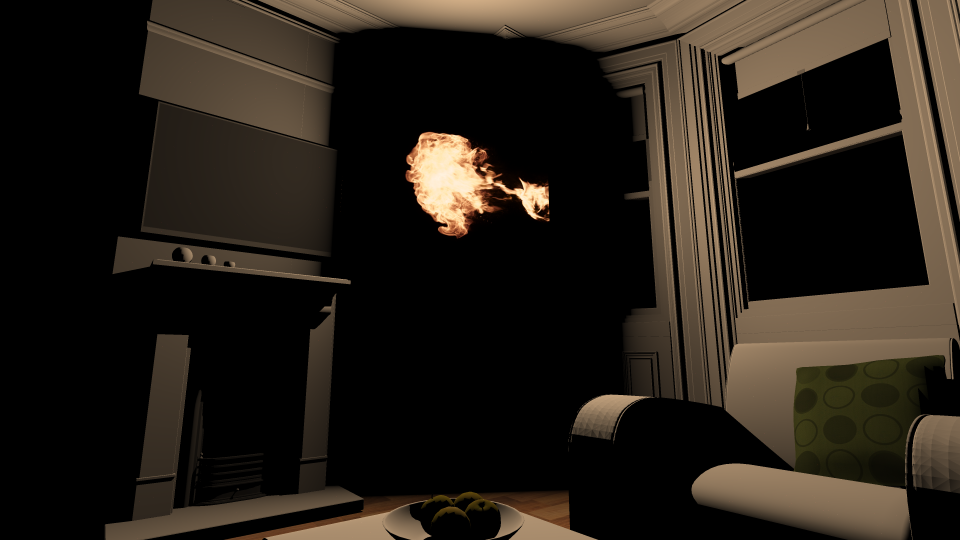}
    & \includegraphics[width=\linewidth]{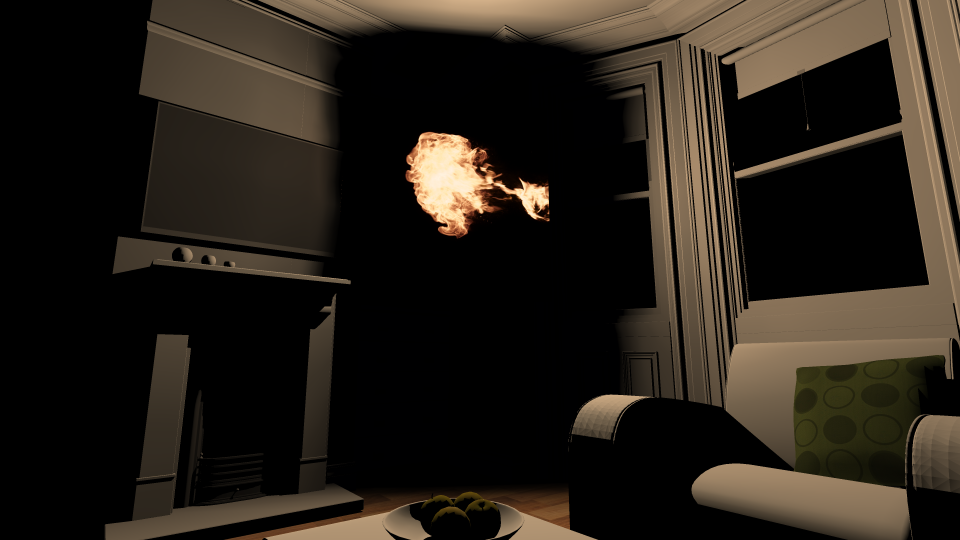}
    & \includegraphics[width=\linewidth]{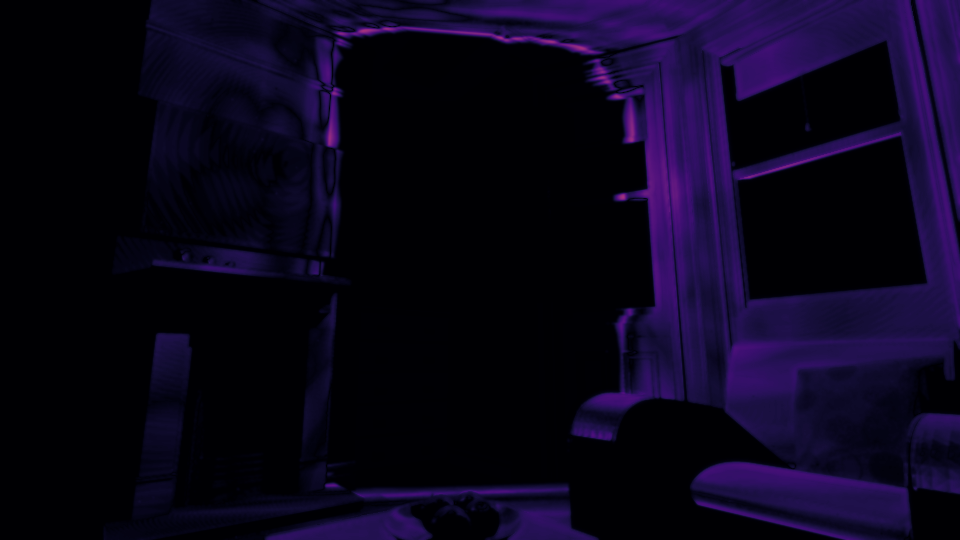} \\[-2pt]
    & & & {\normalsize \FLIP \ ($\downarrow$): 0.0436 } \\[6pt]
    \rotatebox{90}{\small Living Room}
    & \includegraphics[width=\linewidth]{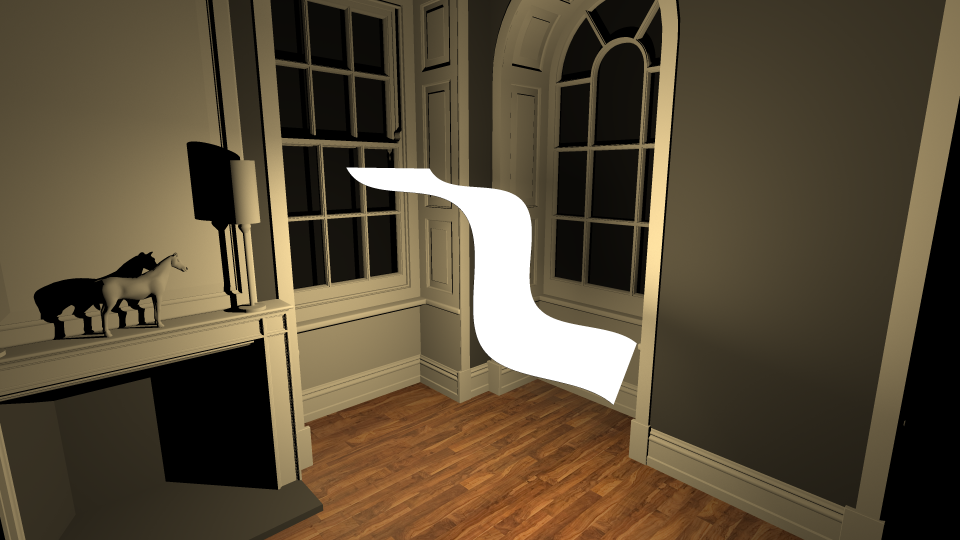}
    & \includegraphics[width=\linewidth]{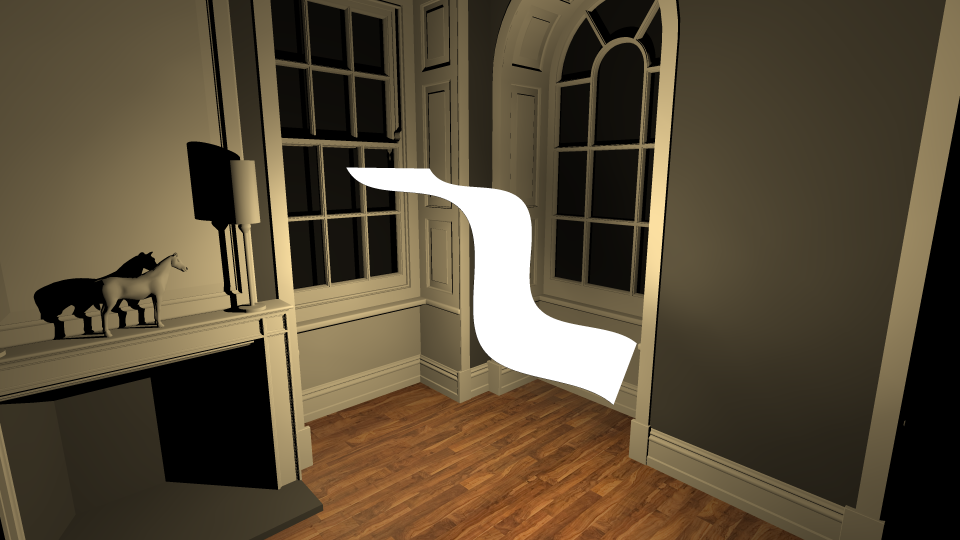}
    & \includegraphics[width=\linewidth]{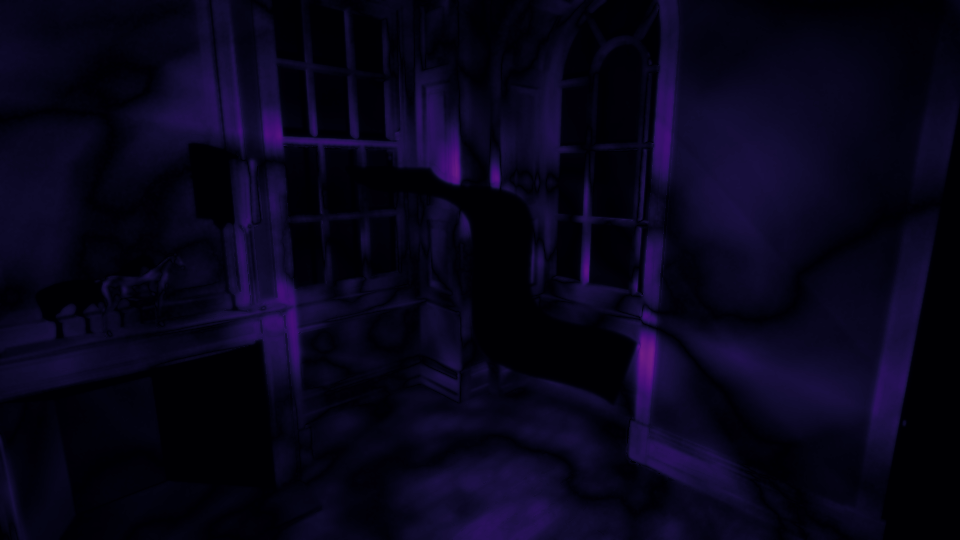} \\[-2pt]
    & & & {\normalsize \FLIP \ ($\downarrow$): 0.0425 } \\
  \end{tabular}
  \vspace{-1em}
  \caption{Comparison of NEF predictions against Monte Carlo references across five scenes with different emitter configurations. \textit{Sponza} contains three distinct emitter types (trained individually), \textit{Bistro} contains six instances of the \textit{Pierced Sphere}  emitter, Neon NEF contains three colored emissive meshes as a single emitter, the \textit{Fire Ball} (in \textit{Whiteroom}), and a deformable sine wave emitter (in \textit{Living Room}) change as a function of a time parameter~\textit{t}. All \FLIP{} maps share a common color scale.}
  \label{fig:full_page_2}
\end{figure*}

\textbf{Dynamic and Deformable Emitters.} 
To support dynamic or deformable emitters, we extend NEF with a time-varying parameter $t$. As with the higher-order emissive effects in the previous section, this only requires that training data collection account for the time-varying state of the emitter. \figref{fig:deformable-emitter} shows 
a textured fireball whose time parameter indexes into frames of a source video. See Figure~\ref{fig:full_page_2} for an emitter with deformable geometry. We encourage the reader to consult the supplemental video for a detailed demonstration of the results.

\subsection{Comparison to Sampling-based Methods} \label{sec:restir}

Both online many-light sampling methods and proxy-based emitters~\cite{heidrich1998canned,velazquez2015complex,zhu2021neural}
require an integration step at runtime. We compare against a representative
of each category: ReSTIR~\cite{bitterli2020spatiotemporal}  and an \textit{oracle} proxy formulation.

\textbf{Many-light Sampling.}  
Unlike ReSTIR, which operates entirely at runtime with no precomputation,
NEF amortizes work into a one-time training phase in exchange for a
constant-time per-pixel inference cost. To make this trade-off concrete,
we construct a Cornell Box scene with 1,000 point lights treated as a single composite emitter: the lights support global rigid transformations, but are fixed relative to each other. A single NEF is trained on this composite.

\begin{figure}[t]
    \centering
    \includegraphics[width=\linewidth]{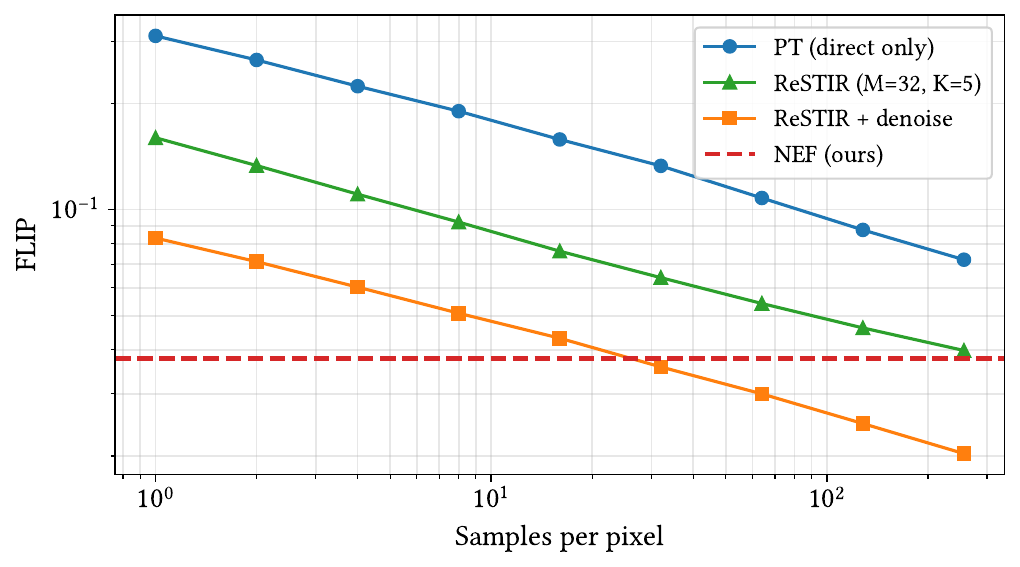}
    \vspace{-2em}
    \caption{Equal-quality convergence on a Cornell box with 1000 point
    lights. NEF (dashed) reaches a fixed error level via a single network
    evaluation per pixel; ReSTIR DI and its denoised variant require
    32 (with denoising) and 256 samples (without) to reach a comparable error, well above
    real-time budgets. Path tracing of direct illumination is included for
    reference. Training cost is amortized
    across frames and is not included in the per-frame timings.}
    \label{fig:restir_convergence}
\end{figure}

We evaluate equal-quality convergence against three baselines: 
path-traced direct illumination, ReSTIR DI ($M{=}32$, $K{=}5$), and ReSTIR DI
paired with an OptiX denoiser. To reach NEF's error level, ReSTIR requires $\sim$256~SPP
without denoising, or $\sim$32~SPP with denoising (\figref{fig:restir_convergence}). While both unbiased sampling baselines eventually surpass NEF at high sample counts, even the 32~SPP denoised case exceeds the typical per-frame budget of real-time pipelines.
However, ReSTIR remains more general: it handles arbitrary per-light motion at runtime, whereas NEF requires the emitter configuration to be fixed at training time. 
Rigid transforms of the entire emitter group and pre-arranged relative motion of the emitters in the composition (via a time parameter, \secref{sec:meshes}) are supported, but arbitrary independent light motion requires retraining.

\textbf{Proxy Emitters.}
\begin{figure}
    \centering
    \setlength{\tabcolsep}{2pt}
    \newlength{\imgw}\setlength{\imgw}{0.43\linewidth}%
    \begin{tabular}{@{}ccc@{}}
        \begin{tikzpicture}[baseline=(n.center)]
            \node[inner sep=0pt] (n) {\includegraphics[width=\imgw]{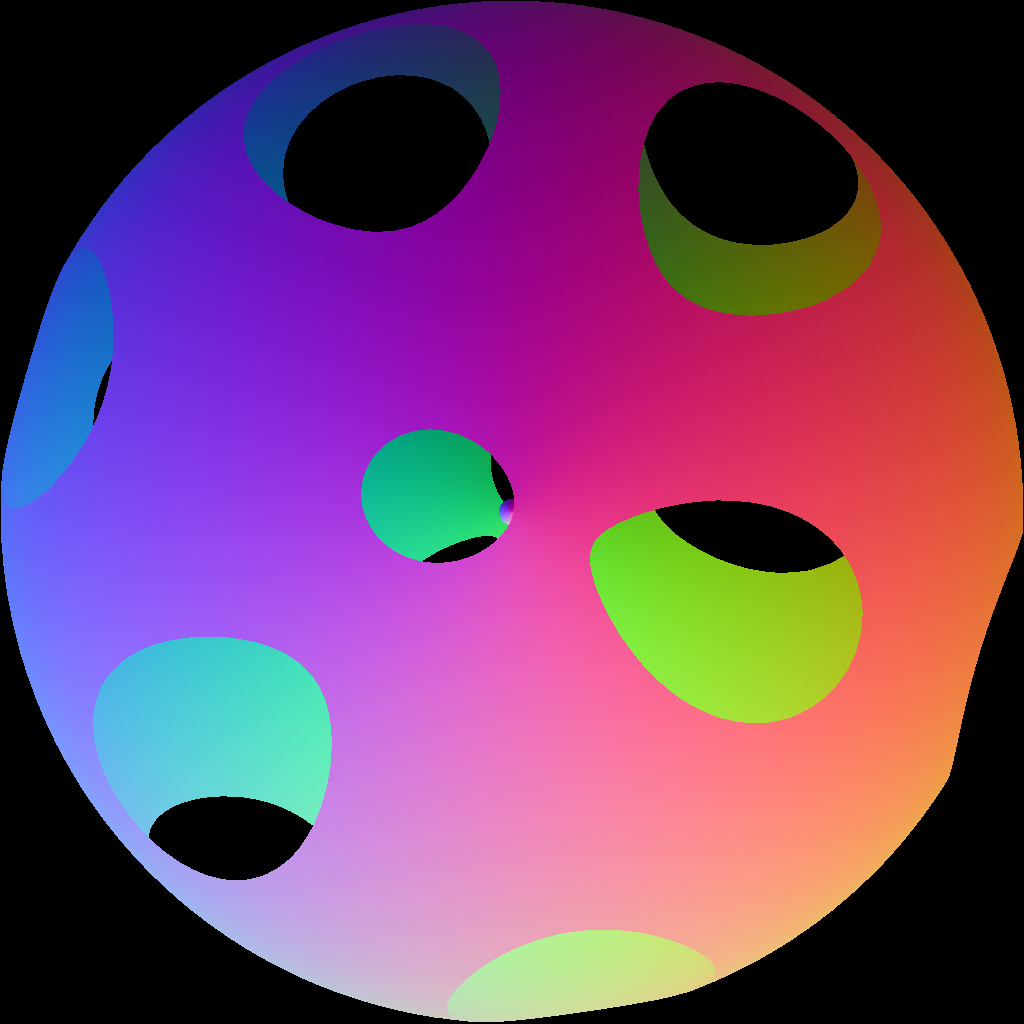}};
        \end{tikzpicture} &
        \begin{tikzpicture}[baseline=(rad.center)]
            \node[inner sep=0pt] (rad) {\includegraphics[width=\imgw]{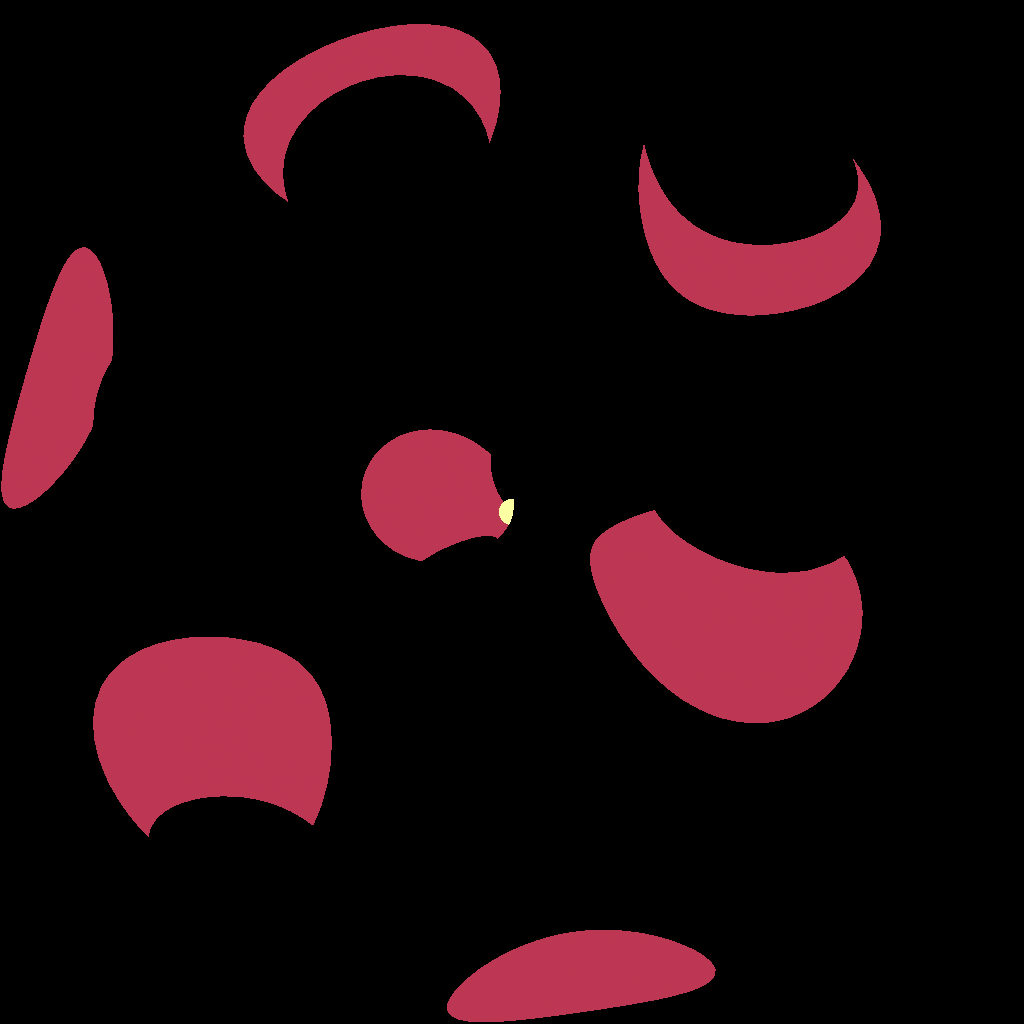}};
            \draw[white, dashed, line width=0.8pt, opacity=0.8, dash phase=1pt]
                (rad.center) circle[radius=0.49
                7\imgw];
        \end{tikzpicture} &
        \begin{tikzpicture}[baseline=(strip.center)]
            \node[inner sep=0pt, draw, thin] (strip) {\includegraphics[height=\imgw]{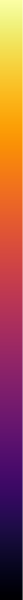}};
            \foreach \v/\lbl in {-0.5/0.0, -0.3/0.2, -0.1/0.4, 0.1/0.6, 0.3/0.8, 0.5/1.0} {
                \node[right, font=\small, inner sep=1pt] at ([yshift=\v*\imgw]strip.east) {\lbl};
            }
        \end{tikzpicture} \\
        \small (a) Surface normals & \small (b) Incident radiance $L_i(\omega)$ & \\[4pt]
    \end{tabular}
    \includegraphics[width=\linewidth]{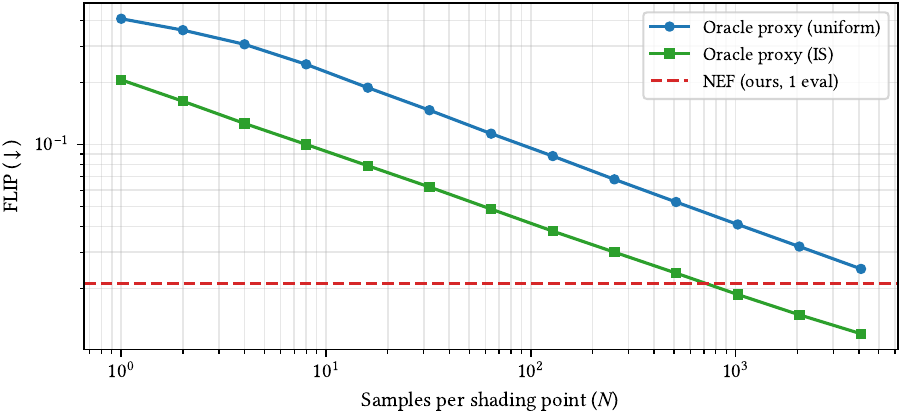}
    \vspace{-2em}
    \caption{%
    Top: the proxy sampling domain as seen from a receiver point,
    showing (a)~surface normals of the pierced-sphere luminaire and
    (b)~the incident radiance field $L_i(\omega)$ that must be integrated.
    The dashed circle marks the proxy sphere extent.
    Bottom: even importance sampling~\cite{zhu2021neural} the proxy,
    requires $734$ samples per shading point
    to match a single NEF evaluation (dashed red). Uniform cone sampling requires ten times more samples.}
    \label{fig:proxy_convergence}
\end{figure}
Proxy representations---whether using 4D light-field slabs~\cite{heidrich1998canned}, a radiance volume with virtual point lights~\cite{velazquez2015complex}, or neural fields~\cite{zhu2021neural}---gather radiance by sampling and integrating a proxy representation.
We compare NEF against an \emph{oracle} proxy: an idealized proxy whose
near-field queries return a noise-free reference and whose evaluation
cost is treated as zero. Any sampling-based proxy method is bounded
from below by this oracle, so the comparison reveals the minimum number
of proxy evaluations required to match NEF's pre-integrated estimate,
regardless of how the proxy itself is represented.

As shown in \figref{fig:proxy_convergence}, for the \textit{Pierced Sphere} scene (\figref{fig:emissive_meshes}),
the oracle proxy requires approximately 734 samples per pixel to
match the accuracy of a single NEF query when combined with a learned importance sampling routine from~\cite{zhu2021neural}.
Each of those samples
still requires a BSDF evaluation at the shading point: at $1920\times1080$ resolution, the 734 GGX evaluations alone cost $2.9$~ms per frame in an optimized
CUDA kernel. This already exceeds NEF's total inference
cost of $2.5$~ms, before any of the proxy queries or importance sampling are accounted
for.
Moreover, the cost of those proxy queries is not negligible in practice. As a data point, a half-precision \texttt{torch.compile}'d forward
pass of the evaluation network from~\cite{zhu2021neural} takes $103$~ms
per emitter sample per full-HD frame, and $49$~ms to output the importance sampling distributions on the same hardware as our NEF timings.

\section{Discussion}

\begin{figure}
    \centering
    \setlength{\tabcolsep}{1.5pt}
    \renewcommand{\arraystretch}{0.6}
    \newcommand{\tilewidth}{0.24\linewidth}
    \begin{tabular}{cccc}
        {\footnotesize Reference} & {\footnotesize OptiX 1 spp} & {\footnotesize OptiX 2 spp} & {\footnotesize OptiX 4 spp} \\[2pt]
        \includegraphics[width=\tilewidth]{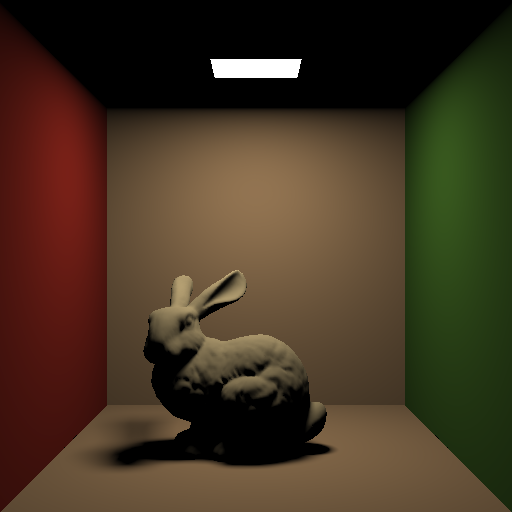} &
        \includegraphics[width=\tilewidth]{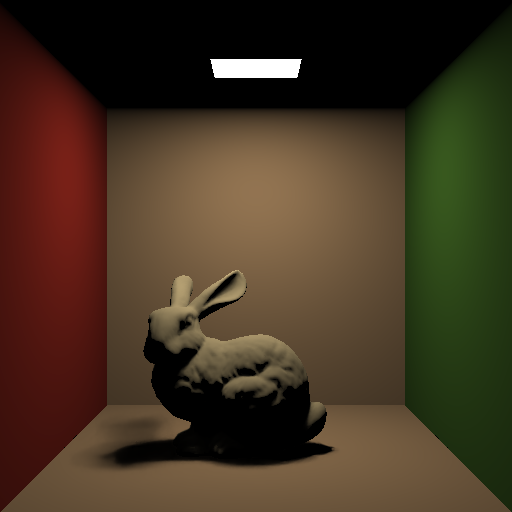} &
        \includegraphics[width=\tilewidth]{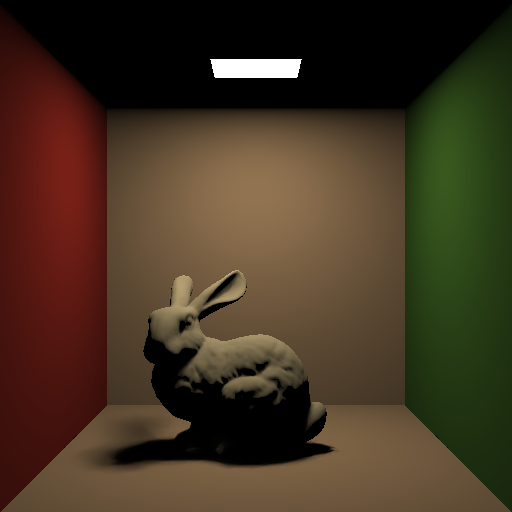} &
        \includegraphics[width=\tilewidth]{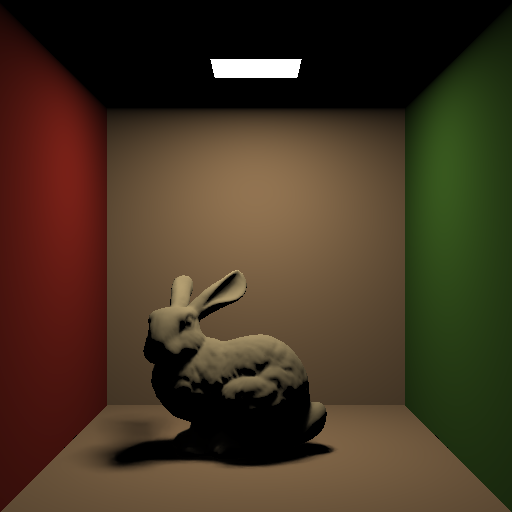} \\[2pt]
        &
        \includegraphics[width=\tilewidth]{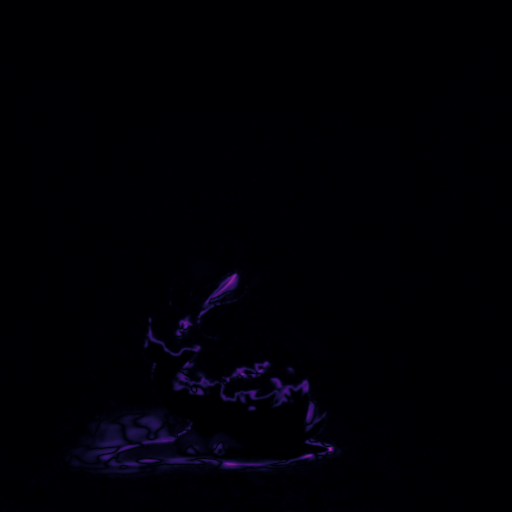} &
        \includegraphics[width=\tilewidth]{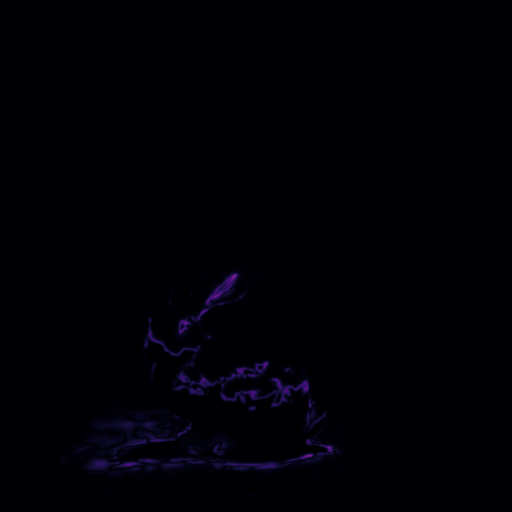} &
        \includegraphics[width=\tilewidth]{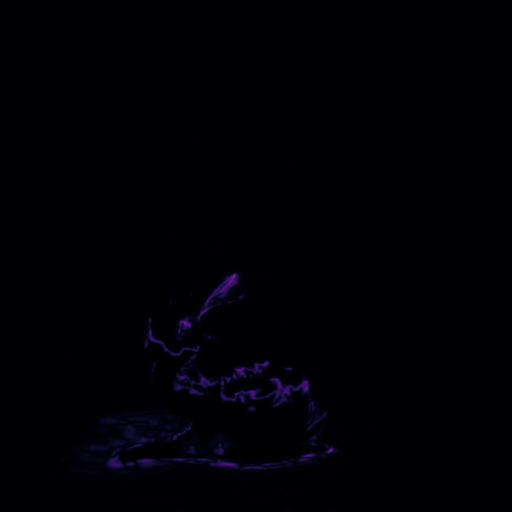} \\[2pt]
        & {\footnotesize \FLIP: 0.0071} & {\footnotesize \FLIP: 0.0062} & {\footnotesize \FLIP: 0.0057} \\
    \end{tabular}
    \vspace{-1em}
    \caption{Ratio estimator~\cite{heitz2018combining} with the OptiX denoiser. The \FLIP{} error map is in range $[0,1]$. Even at 1 shadow ray per pixel, the
    combination of NEF's unshadowed illumination and denoised visibility produces perceptually accurate soft shadows.}
    \label{fig:ratio-estimator}
\end{figure}

\textbf{Visibility and soft shadows.}
Similarly to analytical methods and proxy emitters, NEF only captures the unoccluded emission field.
In practice, this is well-addressed by combining NEF shading with a small budget of path-traced visibility samples and a denoiser that are used as a ratio estimator~\cite{heitz2018combining}. This approach yields plausible soft shadows while preserving the noise-free direct illumination produced by the NEF. 
Even 1 shadow ray per pixel yields perceptually accurate soft shadows (see \figref{fig:ratio-estimator}), with further samples producing diminishing returns. This makes NEF compatible with standard real-time visibility pipelines at minimal additional cost.

\textbf{Storage and training.}
All results in this paper were generated using a single NEF architecture totaling 15~MB (or 30~MB for the glossy variant) per emitter. For simpler emitters, this is often excessive and can be reduced (see supplemental). Training is a one-time offline cost, ranging from approximately one minute for simple planar emitters to three hours for emitters with internal interreflections, depending primarily on the cost of generating path-traced training samples. Crucially, this cost is incurred once per emitter and can be amortized across multiple scenes.

\textbf{Multiple emitters in a scene.}
Currently, a scene with $N$ emitters requires evaluating $N$ independent NEFs and summing their contributions (unless emitters are close enough to be represented jointly by a single NEF), scaling linearly with $N$. We see two avenues to mitigate this. First, a shared decoding MLP could learn to disentangle latents from multiple emitters; however a naive sum-then-decode scheme retains a linear cost for the latent fetching. Second, future work could combine NEF with lighting hierarchies~\cite{walter2005lightcuts, moreau2019dynamic} to achieve sub-linear scaling.

\textbf{Sampling training data.} We generate training targets by path tracing the emitter assembly with next-event estimation. For emitters with housing geometry more intricate than those presented in this work---heavily enclosing the emissive surface or admitting light only through narrow apertures---this strategy can produce a sparse training signal: most paths fail to reach the emitter and contribute zero, leaving the network to learn from a small fraction of informative samples. As shown in recent work~\cite{figueiredo2026purelight}, light tracing~\cite{arvo1986backward} can reduce the number of zero-radiance contributions. Orthogonally, when the emissive field around the emitter exhibits high-frequency variation, adaptive sampling the training loss~\cite{diolatzis2022active} could training data on the regions that need it most.

\section{Conclusion}
We presented Neural Emission Fields, a pre-integrated neural representation of direct illumination from non-trivial emitters. By moving the receiver-side integral into an offline training phase, NEF reduces shading to a single network evaluation per pixel, at a cost independent of the emitter's geometric complexity. 
The same formulation handles a wide range of lighting scenarios---including spatially varying emission, internal interreflections, and deformable emitters---within a single, unified training procedure; trained models then act as portable lighting assets that can be placed in any scene under rigid transforms and uniform scaling. 
We hope NEF proves a useful step toward treating expensive direct illumination as a bakeable, reusable component of real-time rendering pipelines.

\clearpage

\clearpage

\bibliographystyle{eg-alpha-doi} 
\bibliography{bibliography}

\clearpage
\appendix

\section{Linear Scaling of Analytical Methods}
\begin{figure}[h!]
    \centering
    \includegraphics[width=\linewidth]{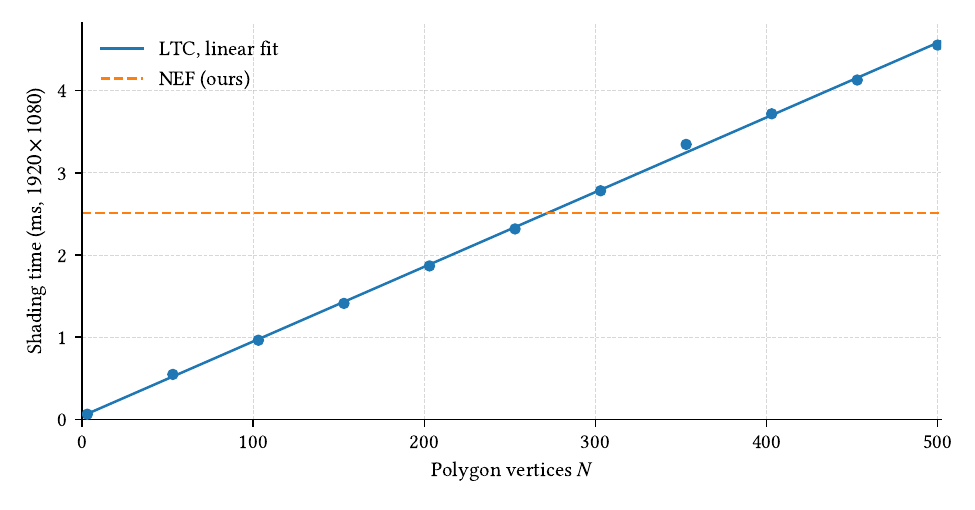}
    \vspace{-2.5em}
    \caption{
Linear Scaling vs. Constant-Time Inference. Shading cost (1080p) for analytical methods scales linearly with vertex count $N$. Our method (NEF) exhibits constant-time performance, outperforming the LTC reference~\cite{ltc2016} beyond the crossover point of 267 vertices.
}
    \label{fig:analytical_scaling}
\end{figure}

As shown in \figref{fig:analytical_scaling}, the computational cost of analytical methods like LTC~\cite{ltc2016} scales linearly with the number of emitter vertices. In contrast, NEF inference time remains constant for a given emitter geometry, regardless of the density of its tessellation. 
For glossy receivers, our method reaches a performance crossover at 267 emitter vertices. Notably, NEF maintains a constant 2.5ms shading time even for highly complex emitters like the \textit{Stanford Dragon} (436k vertices), where analytical methods would be computationally prohibitive.

\section{Representation Parameters}

We use a single set of hash grid parameters for all encoded inputs. The GlossyNEF architecture shares the positional encoding between the diffuse and glossy head. The multiresolution hash table has eight levels, each level increases in size with a rate of $\sqrt{2}$ per level, the corresponding latent vector consists of four floats, and the hash table uses $2^{19}$ entries. The additional inputs to the neural network, and the overall memory budget in half-precision, are summarized in Table~\ref{tab:architecture}.

\begin{table}[h!]
  \centering
  \caption{Network architecture for the diffuse and glossy NEF variants.
  The glossy model uses a two-head architecture: a diffuse head with the same
  architecture predicts irradiance, while the glossy head predicts the glossy component.}
  \label{tab:architecture}
  \begin{tabular}{lll}
    \toprule
    Property & Diffuse NEF & Glossy NEF \\
    \midrule
    Position encoding & Hash grid on $x$ & Hash grid on $x$\\
    Direction encoding & --- & Hash grid on $\omega_r$ \\
    Normal & $n \in [-1,1]^3$  & $n \in [-1,1]^3$ \\
    Roughness & --- & $\alpha \in [0.1,1]$ \\
    Hidden layers & $4 \times 128$, ReLU & $4 \times 128$, ReLU \\
    Output & 3 (RGB), linear & 3 (RGB), linear \\
    Storage (FP16) & 15.22 MB & 30.46 MB \\
    \bottomrule
  \end{tabular}
\end{table}

\section{Architecture Ablation}
We ablate several architecture choices (see Table~\ref{tab:ablation}) of our NEF by training a GlossyNEF on the same \textit{Quad} emitter with the same number of training iterations. 

Applying the multiresolution hash encoding on the outgoing direction as opposed to the reflected direction, which is a common encoding choice for neural representations, reduces quality across all roughness values $\alpha$.

Splitting the network into a diffuse and glossy head similarly performs better than unifying into a single head.

The original NEF architecture shares the positional encoding between the diffuse and glossy head. Decoupling this multiresolution hash table (i.e. diffuse and glossy each use a different one) yields no quality improvement, although it adds more capacity and extra runtime.

\begin{table}[h!]
    \centering
    \caption{Ablation of the GlossyNEF architecture.
    We report \FLIP{} ($\downarrow$) at three roughness levels.
    The baseline uses two output heads with hash-grid encodings on both
    position~$x$ and the reflected view direction~$\boldsymbol{\omega}_r$.
    The variant marked $2\!\times$ doubles the capacity of the
    remaining component to control for parameter count.}
    \label{tab:ablation-glossy}
    \small
    \begin{tabular}{@{}l ccc@{}}
        \toprule
        & $\alpha=0.1$ & $\alpha=0.3$ & $\alpha=0.8$ \\
        \midrule
        Full (ours)
        & \textbf{0.0216} & \textbf{0.0230} & \textbf{0.0172} \\
        \addlinespace
        hash($\boldsymbol{\omega}_o$) instead of hash($\boldsymbol{\omega}_r$)
        & 0.0299 & 0.0269 & 0.0222 \\
        \addlinespace
        Single head
        & 0.0359 & 0.0302 & 0.0249 \\
        Single head,\; $2\!\times$ MLP
        & 0.0324 & 0.0252 & 0.0204 \\
        \addlinespace
        Decoupled hash($\mathbf{x}$)
        & 0.0246 & 0.0240 & 0.0195 \\
        \bottomrule
    \end{tabular}
    \label{tab:ablation}
\end{table}

\section{Training Budget}
Table~\ref{tab:training-budget} reports the training cost for each NEF in the paper, measured on a single RTX 4090 GPU. 
All timings include the cost of generating path-traced radiance targets on-the-fly, which dominates iteration time for the more complex emitters. Iterations and per-emitter training time scale with the difficulty of the underlying radiance field: planar emitters with uniform emission converge in minutes, bare emissive meshes and spatially varying emitters in under fifteen minutes, while configurations involving internal interreflections or temporal variation require several hours, both because the radiance field itself is more complex and because each path-traced target costs more to generate.

\begin{table}[h!]
    \centering
    \caption{Training budget per NEF on a single RTX 4090 GPU. 
    Sample count is the total number of $(\tilde{x}, \tilde{n}, \tilde{\omega}_o, \alpha)$ queries seen during training (iterations $\times$ batch size $2^{16}$, expressed in millions or billions).}
    \label{tab:training-budget}
    \begin{tabular}{l r r r r}
      \toprule
      Configuration & Iterations & it/s & Training time & Samples \\
      \midrule
      quad\_diffuse & 10,000 & 165 & 1m 0s & 655.4M \\
      dragon & 100,000 & 135 & 12m 22s & 6.6B \\
      pierced\_sphere & 100,000 & 196 & 8m 29s & 6.6B \\
      neon NEF & 100,000 & 142 & 11m 35s & 6.6B \\
      dynamic\_flag & 250,000 & 86 & 48m 20s & 16.4B \\
      fire & 250,000 & 36 & 1h 55m & 16.4B \\
      pierced\_interreflect & 500,000 & 52 & 2h 41m & 32.8B \\
      dragon\_interreflect & 500,000 & 49 & 2h 51m & 32.8B \\
      quad\_glossy & 500,000 & 75 & 1h 51m & 32.8B \\
      \bottomrule
    \end{tabular}
\end{table}

\section{Overspent Capacity} \label{sec:overspent}
\begin{figure}
    \centering
    \includegraphics[width=0.8\linewidth]{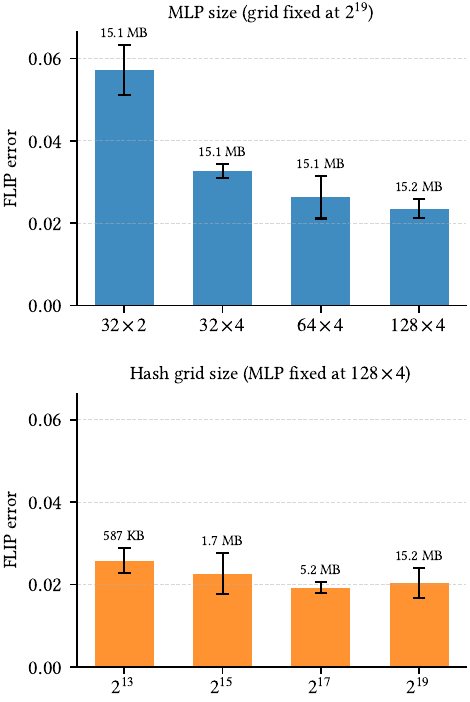}
    \caption{Capacity ablation on a \textit{Quad} emitter. Reducing the multiresolution hash encoding shrinks the footprint substantially with little impact on error, while reducing the MLP causes a rapid quality drop. For simple emissive fields, the default architecture is significantly oversized.}
    \label{fig:smaller_architecture}
\end{figure}

The default NEF architecture is sized for the most demanding emitter in our set (the \textit{Pierced Sphere}) and is excessive for simpler emissive fields. We ablate two capacity axes---the MLP and the multiresolution hash encoding---on a simple \textit{Quad} emitter (Figure~\ref{fig:smaller_architecture}). The two axes have different effects: reducing the MLP size quickly degrades quality, as the network loses the capacity to decode the encoded latents. Shrinking the hash encoding, in contrast, yields the largest reduction in memory footprint at negligible cost in error, since most of the parameter budget lives in the hash tables and the \textit{Quad}'s emissive field does not need the default spatial resolution. In practice, simple emitters can therefore be deployed with a substantially smaller hash table while keeping the MLP near its default size.

\section{Low-roughness Materials}
\begin{figure}
    \centering
    \begin{tabular}{@{}c@{\hspace{4pt}}c@{}}
        \includegraphics[width=0.48\linewidth]{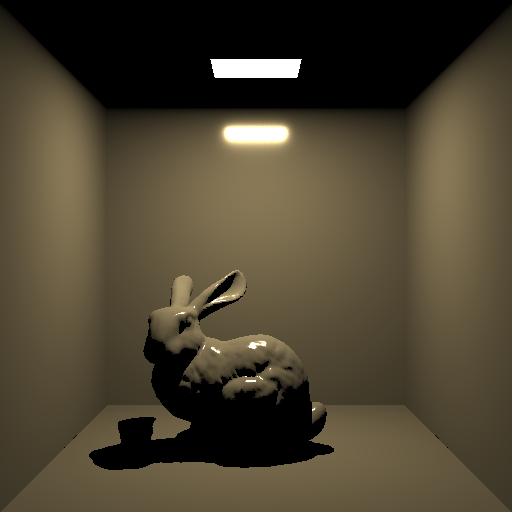} &
        \includegraphics[width=0.48\linewidth]{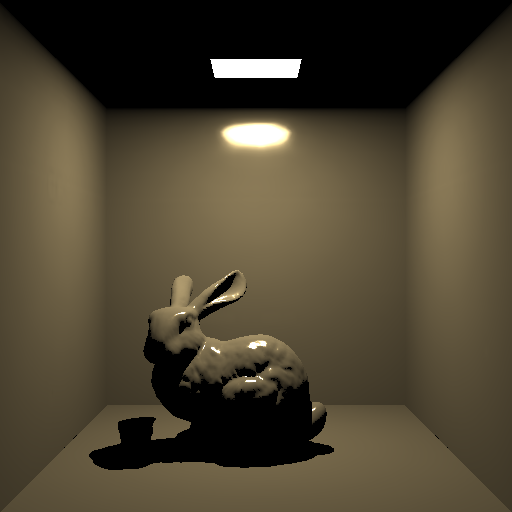} \\[2pt]
        {\small Reference ($\alpha=0.01$)} & {\small NEF ($\alpha=0.01$)} \\
    \end{tabular}

    \vspace{6pt}
    \includegraphics[width=0.95\linewidth]{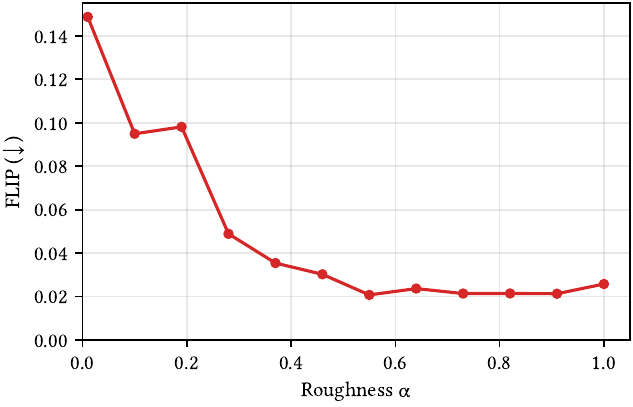}
    \caption{Roughness limitation. As surface roughness $\alpha \to 0$ the specular highlight sharpens beyond the network's representational capacity, increasing perceptual error (\FLIP) in the highlight region. NEF was trained on the extended range of $\alpha \in [0.01, 1]$ for this figure.}
    \label{fig:roughness-limit}
\end{figure}

While there are no theoretical constraints preventing NEF from being trained on low-roughness receivers, practical results indicate a bottleneck as $\alpha \to 0$, as illustrated in \figref{fig:roughness-limit}. This degradation occurs because the specular highlight sharpens into a high-frequency signal that exceeds the network’s representational capacity. Consequently, while training is technically feasible across the entire roughness spectrum, the model struggles to resolve the near-singular gradients of mirror-like reflections, leading to the observed increase in approximation error in the highlight regions.

\section{Impact of Training Data Noise}

\begin{figure}
    \centering
    \setlength{\tabcolsep}{1pt}
    \begin{tabular}{@{}ccc@{}}
        \small $S = 1$ & \small $S = 4$ & \small $S = 16$ \\[2pt]
        \includegraphics[width=0.32\linewidth]{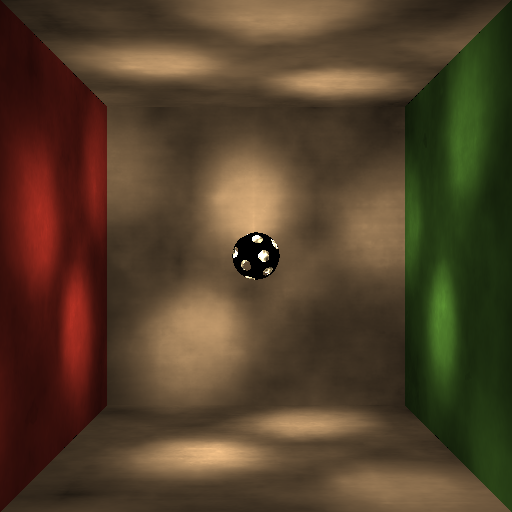} &
        \includegraphics[width=0.32\linewidth]{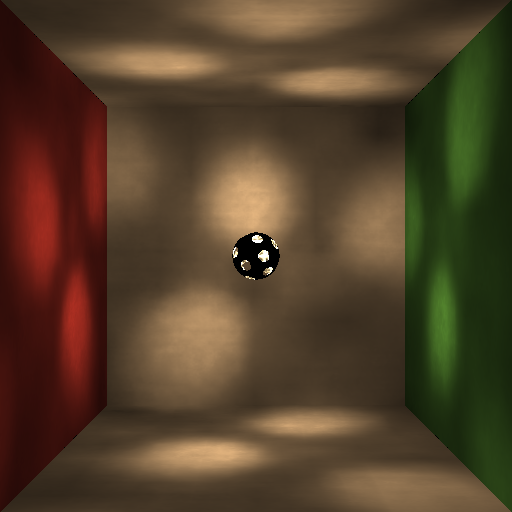} &
        \includegraphics[width=0.32\linewidth]{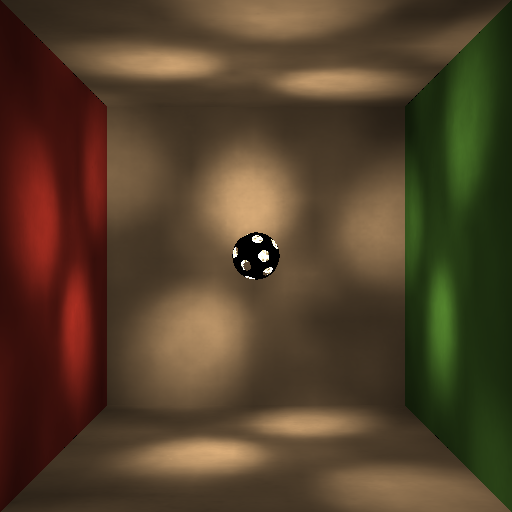} \\[1pt]
        \footnotesize \FLIP\ 0.082 &
        \footnotesize \FLIP\ 0.061 &
        \footnotesize \FLIP\ 0.046 \\[6pt]
        \small $S = 64$ & \small $S = 256$ & \small Reference \\[2pt]
        \includegraphics[width=0.32\linewidth]{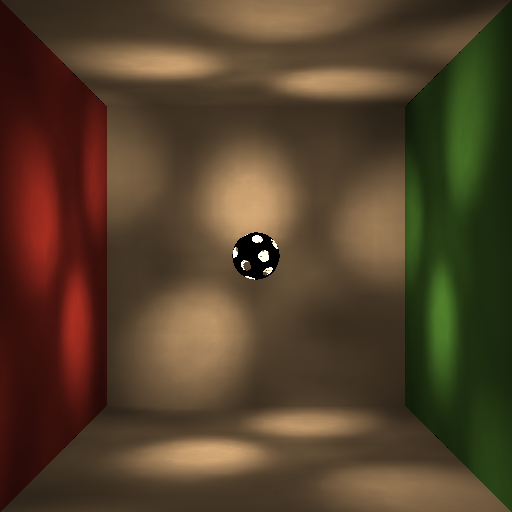} &
        \includegraphics[width=0.32\linewidth]{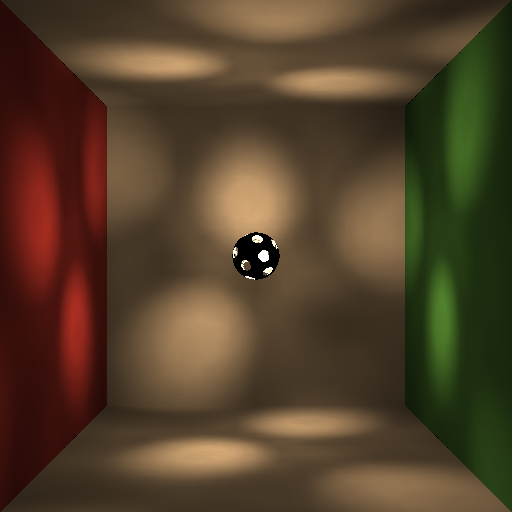} &
        \includegraphics[width=0.32\linewidth]{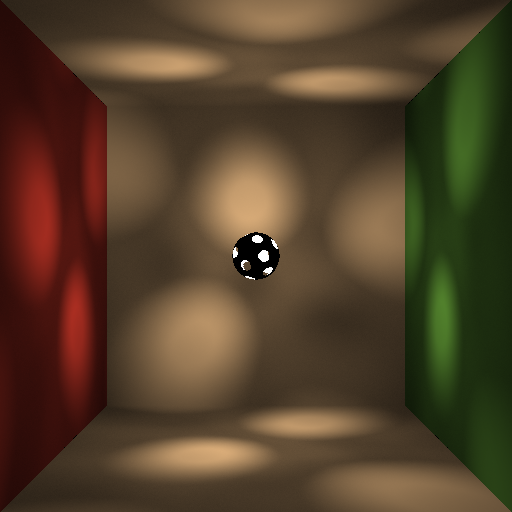} \\[1pt]
        \footnotesize \FLIP\ 0.036 &
        \footnotesize \FLIP\ 0.029 & \\[6pt]
    \end{tabular}

    \includegraphics[width=0.98\linewidth]{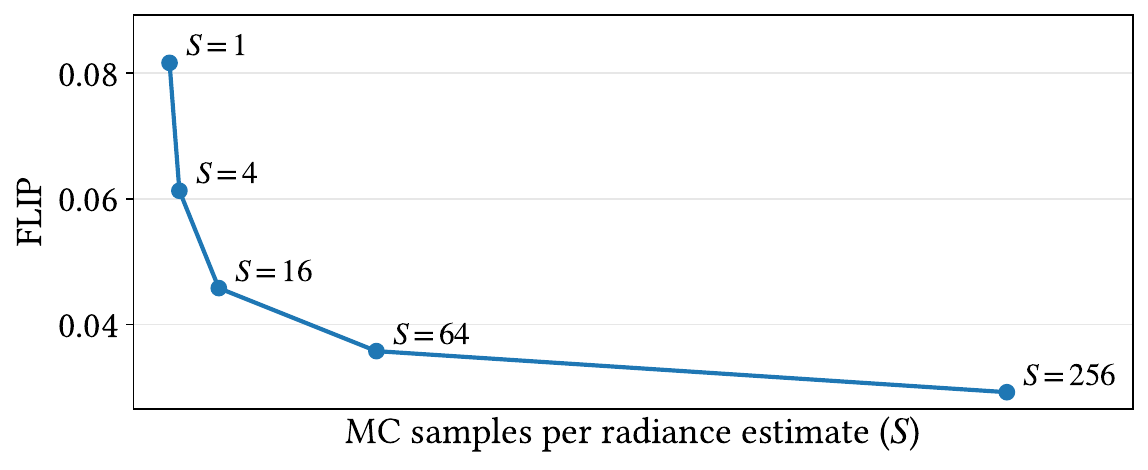}
    \caption{Ablation on the number of MC samples $S$ per training radiance estimate.
    Pierced-sphere emitter, intermediate training result of 50\,k training iterations.
    Higher $S$ reduces training noise and improves the learned field with diminishing returns. }
    \label{fig:ablation_mc}
\end{figure}

The training targets used throughout this paper are generated via path tracing with Monte Carlo integration, leaving residual variance in the supervision signal. The magnitude of this variance depends on the emitter: emitters that are hard to sample yield noisier targets. The \textit{Pierced Sphere} is a particularly challenging case, as its internal interreflections must also be captured during training, making it one of the noisiest emitters in our set. In Figure~\ref{fig:ablation_mc} we ablate the number of Monte Carlo samples used per training estimate and observe diminishing returns beyond 64 samples. We therefore adopt $S = 64$ for all emitters throughout the paper, noting that this budget could safely be reduced for simpler emitters.

\end{document}